\documentclass[twocolumn,twocolappendix]{aastex701}
\usepackage{bm}
\usepackage{hyperref}
\usepackage{amsmath}
\usepackage{mathtools}
\usepackage{url}

\def\eqref#1{equation~(\ref{#1})}

\newcommand {\nn}{\nonumber}

\newcommand {\e}{\,\mathrm{e}}

\newcommand {\sech}{\,{\rm sech}}
\newcommand {\mi}{\mathrm{i}}

\newcommand {\Myr}{\,{\rm Myr}}
\newcommand {\Gyr}{\,{\rm Gyr}}

\newcommand {\pc}{\,{\rm pc}}
\newcommand {\kpc}{\,{\rm kpc}}

\newcommand {\kpcGyr}{\,{\rm kpc}\,{\rm Gyr}^{-1}}

\newcommand {\kpckpcGyr}{\,{\rm kpc}^2\,{\rm Gyr}^{-1}}
\newcommand {\kmskpc}{\,{\rm km}\,{\rm s}^{-1}\,{\rm kpc}^{-1}}

\newcommand {\K}{\,{\rm K}}

\newcommand {\Msun}{\,{\rm M}_\odot}

\newcommand {\B}[1]{{\boldsymbol{#1}}}

\newcommand {\drm}{\mathrm{d}}

\newcommand {\vx}{{\bm x}}

\newcommand {\vvel}{{\bm v}}
\newcommand {\R}{R}

\newcommand {\Rb}{R_{\rm b}}

\newcommand {\RCR}{R_{\rm CR}}

\newcommand {\Lz}{L_z}

\newcommand {\phib}{\varphi_{\rm b}}

\newcommand {\vJ}{{\bm J}}

\newcommand {\Jr}{J_r}

\newcommand {\Jz}{J_z}

\newcommand {\Jphi}{J_\varphi}

\newcommand {\vJf}{{\bm J}_{\rm f}}
\newcommand {\Jfo}{J_{{\rm f}_1}}
\newcommand {\Jft}{J_{{\rm f}_2}}
\newcommand {\Js}{J_{\rm s}}

\newcommand {\Jsres}{J_{\rm s, res}}

\newcommand {\Jl}{J_\ell}

\newcommand {\jl}{j_\ell}
\newcommand {\jlsep}{j_{\ell,{\rm sep}}}

\newcommand {\vtheta}{{\bm \theta}}

\newcommand {\thetar}{\theta_r}

\newcommand {\thetaz}{\theta_z}
\newcommand {\thetaphi}{\theta_\varphi}

\newcommand {\thetas}{\theta_{\rm s}}

\newcommand {\thetafo}{\theta_{{\rm f}_1}}
\newcommand {\thetaft}{\theta_{{\rm f}_2}}

\newcommand {\thetasres}{\theta_{\rm s,res}}

\newcommand {\vthetaf}{{\bm \theta}_{\rm f}}

\newcommand {\phil}{\phi_\ell}

\newcommand {\Omegap}{\Omega_{\rm p}}

\newcommand {\Omegar}{\Omega_r}
\newcommand {\Omegaz}{\Omega_z}

\newcommand {\Omegaphi}{\Omega_\varphi}

\newcommand {\vOmega}{\mathbf \Omega}

\newcommand {\vN}{{\bm N}}

\newcommand {\Nr}{N_r}

\newcommand {\Nphi}{N_\varphi}
\newcommand {\Nz}{N_z}

\newcommand {\hPhi}{\hat{\Phi}}

\newcommand {\bH}{\bar{H}}
\newcommand {\bHres}{\bar{H}_{\rm res}}
\newcommand {\bHsep}{\bar{H}_{\rm sep}}

\newcommand {\Tl}{T_\ell}

\newcommand {\pd}{{\partial}}

\newcommand {\M}{{\rm M}}

\newcommand {\rh}{r_{\rm h}}

\newcommand {\Rd}{R_{\rm d}}
\newcommand {\rb}{r_{\rm b}}
\newcommand {\rhoh}{\rho_{\rm h}}
\newcommand {\rhod}{\rho_{\rm d}}
\newcommand {\rhob}{\rho_{\rm b}}
\newcommand {\zd}{z_{\rm d}}

\newcommand {\Mh}{M_{\rm h}}
\newcommand {\Md}{M_{\rm d}}

\begin{document}

\title{Tree-ring structure of Galactic bar resonance in $N$-body simulations}

\author[orcid=0000-0002-3445-855X,gname=Rimpei,sname=Chiba]{Rimpei Chiba}
\affiliation{Department of Astronomy, Graduate School of Science, The University of Tokyo, 7-3-1 Hongo, Bunkyo-ku, Tokyo, 113-0033, Japan}
\affiliation{Canadian Institute for Theoretical Astrophysics, University of Toronto, 60 St. George Street, Toronto, ON M5S 3H8, Canada}
\email[show]{rimpei-chiba@g.ecc.u-tokyo.ac.jp}  

\author[orcid=0000-0002-6465-2978,gname=Michiko,sname=Fujii]{Michiko Fujii}
\affiliation{Department of Astronomy, Graduate School of Science, The University of Tokyo, 7-3-1 Hongo, Bunkyo-ku, Tokyo, 113-0033, Japan}
\email{fujii@astron.s.u-tokyo.ac.jp}

\author[orcid=0000-0002-2154-8740,gname=Junichi,sname=Baba]{Junichi Baba}
\affiliation{Amanogawa Galaxy Astronomy Research Center, Graduate School of Science and Engineering, Kagoshima University, 1-21-35 Korimoto, Kagoshima 890-0065, Japan}
\affiliation{Division of Science, National Astronomical Observatory of Japan, Mitaka, Tokyo 181-8588, Japan}
\email{babajn2000@gmail.com}

\author[orcid=0000-0001-5337-0732,gname=John,sname=Dubinski]{John Dubinski}
\affiliation{Canadian Institute for Theoretical Astrophysics, University of Toronto, 60 St. George Street, Toronto, ON M5S 3H8, Canada}
\email{dubinski@cita.utoronto.ca}

\author[orcid=0000-0002-4236-3091,gname=Ralph,sname=Sch\"onrich]{Ralph Sch\"onrich}
\affiliation{Mullard Space Science Laboratory, University College London, Holmbury St. Mary, Dorking, Surrey, RH5 6NT, UK}
\email{r.schoenrich@ucl.ac.uk}

%% Use the \collaboration command to identify collaborations. This command
%% takes an optional argument that is either a number or the word "all"
%% which tells the compiler how many of the authors above the command to
%% show. For example "\collaboration[all]{(DELVE Collaboration)}" wil include
%% all the authors above this command.
%%
%% Mark off the abstract in the ``abstract'' environment. 
\begin{abstract}

We study the structure and evolution of the galactic bar's resonant phase-space in self-consistent $N$-body simulations of the Milky Way, with and without perturbations from the Sagittarius dwarf galaxy. In an idealized disk evolution model in which stars are perturbed solely by a bar that spins down due to dynamical friction against the dark matter halo, it is predicted that stars trapped in the bar's corotation resonance form a characteristic `tree-ring' structure in phase space: as the resonance expands in volume while sweeping outwards, it sequentially captures surrounding stars at its surface, such that stars captured earlier in the inner disk are found preferentially near the core of the resonance. However, it has not been clear whether such a structure persists in a more realistic galactic disk subject to a variety of time-dependent perturbations, in particular those by spiral arms and passing satellite galaxies. This paper demonstrates that the predicted tree-ring structure indeed emerges in a realistic noisy environment using self-consistent $N$-body simulations. Despite the presence of spiral arms, encounters with the Sagittarius dwarf galaxy, as well as fluctuations in the bar's pattern speed, and not least numerical noise---all of which drive stellar diffusion in phase space---the tree-ring structure remains well-preserved in the slow angle-action space. Our results demonstrate that the tree-ring structure of the bar's resonance is a robust signal of the bar's spin-down and hence its discovery in the Milky Way implies the existence of a dark matter halo that removed angular momentum from the bar.

\end{abstract}

%% Keywords should appear after the \end{abstract} command. 
%% The AAS Journals now uses Unified Astronomy Thesaurus (UAT) concepts:
%% https://astrothesaurus.org
%% You will be asked to selected these concepts during the submission process
%% but this old "keyword" functionality is maintained in case authors want
%% to include these concepts in their preprints.
%%
%% You can use the \uat command to link your UAT concepts back its source.

%% From the front matter, we move on to the body of the paper.
%% Sections are demarcated by \section and \subsection, respectively.
%% Observe the use of the LaTeX \label
%% command after the \subsection to give a symbolic KEY to the
%% subsection for cross-referencing in a \ref command.
%% You can use LaTeX's \ref and \label commands to keep track of
%% cross-references to sections, equations, tables, and figures.
%% That way, if you change the order of any elements, LaTeX will
%% automatically renumber them.

\section{Introduction}
\label{sec:Introduction}

% Introduction on the bar's secular evolution
The Milky Way possesses a central Galactic bar, a structure observed in more than two-thirds of local disk galaxies \citep[e.g.][]{Erwin2018dependence}. Bars serve as sensitive probes of dark matter because their evolution depends critically on its existence: bars lose angular momentum to dark matter by dynamical friction \citep[e.g.][]{Sellwood1980Galaxy,weinberg1985evolution,Athanassoula1996Evolution}, while gaining it from cold gas in the inner disk \citep[e.g.][]{Friedli1993Secular,Berentzen2007Gas,VillaVargas2010Gas,Beane2023StellarBars,Kwak2026SMUGGLE}, with the former typically outweighing the latter, thereby causing bars to spin down over time \citep{Semczuk2024Pattern}\footnote{See, however, \cite{Merrow2026What}, who find accelerating bars in baryon-dominated galaxies in the Auriga simulations.}. As bars slow, they typically also grow in both strength and length \citep[e.g.][]{Athanassoula2002BarGrowth,Martinez2006Evolution,aumer2015origin}. In contrast, bars evolved in modified gravity without dark matter retain an almost constant pattern speed and amplitude \citep{Tiret2007MOND,Ghafourian2020Modified}, or even spin up and weaken in the presence of gas \citep{Nagesh2023Simulations}. Identifying whether---or how rapidly---bars spin down in real galaxies thus provides key constraints on the existence and nature of dark matter.

% Tree-ring structure
In \cite{Chiba2020ResonanceSweeping} and \cite{Chiba2021TreeRing}, we demonstrated that the spin-down of the bar leaves an observable imprint on the phase space of the bar's resonance. When the bar slows, its resonance grows in phase-space volume as it sweeps toward larger radii, thereby capturing a fraction of stars along its path. Because the trapped stars librate around the resonance while adiabatically conserving the enclosed phase-space area, the resonance develops just like tree rings, where stars captured earlier in the inner disk occupy the core of the resonance, while those captured later at larger radii are found near the surface (separatrix) of the resonance. This suggests that the stellar metallicity, which correlates with the galactocentric radius at which stars are born, should increase monotonically toward the resonance center. \cite{Chiba2021TreeRing} found this trend in the Milky Way using the Gaia data and interpreted it as evidence for the bar's spin-down.

% Other neglected perturbations
However, the tree-ring structure of the bar's resonance has so far been demonstrated only in idealized models where the stellar disk is perturbed only by a slowing bar. Stars in real galaxies experience a range of additional perturbations, most notably from galactic spiral arms and passing satellite galaxies. It has long been recognized that spiral arms cause significant changes in the angular momenta of stars in the disk \cite[e.g.][]{lynden1972generating,Carlberg1985Dynamical,Sellwood2002radial,Minchev2011Radial,Roskar2012Radial}. For example, \cite{Hunt2018Transient,Hunt2019signature} showed that perturbations from a series of transient winding spiral arms can significantly smear signatures of the bar's resonances, making even their locations difficult to identify. 

Passing satellite galaxies and dark subhalos may also produce significant in-plane perturbations in the disk through tidal forcing \citep[e.g.][]{Toomre1972GalacticBridges,Noguchi1987Close,Oh2008Physical}. Among the satellites in the Milky Way (MW), the Sagittarius dwarf galaxy (Sgr) is believed to have caused by far the largest perturbations in the disk over the past $\sim8$ Gyr \citep{Banik2022ComprehensiveI,Banik2023ComprehensiveII}. The MW--Sgr interaction has been explored extensively \citep[e.g.][]{Binney2018GaiaPhaseSpiral,Laporte2019Footprints,Vasiliev2021Tango,BlandHawthorn2021GalacticSeismology,Hunt2021Resolving,Bennett2022Exploring,Asano2025Ripples}, motivated in large part by the Gaia's recent discovery of the vertical phase-space spiral \citep{Antoja2018Nature}. Yet the impact of Sagittarius on the phase-space distribution of stars trapped in the bar's resonance remains largely unexplored.

% Theoretical prediction
The general expectation is that these additional perturbations cause trapped stars to diffuse through phase space. Stellar diffusion can in fact be driven by a multitude of processes, including encounters with giant molecular clouds \citep[e.g.][]{Spitzer1951Possible,Spitzer1953Possible,Jenkins1990Spiral,Aumer2016Age,Fujimoto2023Efficient}, fluctuations in the bar's pattern speed \citep{Wu2016TimeDependent}, as well as from particle shot noise inherent to $N$-body simulations \citep[e.g.][]{Weinberg2007BarHaloInteraction,Weinberg2007BarHaloInteractionII,Fouvry2015Selfgravity,Ludlow2021Spurious,Wilkinson2023impact}. Using a simple kinetic model with a steadily rotating bar, \cite{Hamilton2023BarResonanceWithDiffusion} recently demonstrated that such diffusion constantly streams stars into and out of the bar's resonance, resulting in a steady-state distribution that remains inhomogeneous along the trajectories of libration, i.e. along the tree-rings.

% Aim of paper
The focus of this paper is to numerically investigate how the phase-space distribution of stars trapped in the bar's resonance evolves in a realistic, noisy environment. To this end, we run fully self-consistent $N$-body simulations of a Milky Way-like galaxy, both with and without the presence of a Sagittarius-like dwarf galaxy. Our goal is to identify the predicted tree-ring structure of the bar's resonance and to assess (i) whether the structure is uniquely associated with the bar's slow-down and (ii) whether it survives in the presence of both internal and external perturbations.

% Structure of paper
The remaining sections are structured as follows. Section~\ref{sec:method} describes the details of the simulations, and Section~\ref{sec:results} presents the results, first identifying structures unique to the bar's spin-down, and later exploring their robustness against external perturbations. Section~\ref{sec:discussion} discusses the limitations of our models, and Section~\ref{sec:Summary} summarizes the results and outlines future directions.

%%%%%%%%%%%%%%%%%%%%%%%%%%%%%%%%%%%%%%%%%%%%%%%%%%%%%%%%%%%%%%%%%%%%%%%%%%%%%%%%%%%%%%%%%%%%%%%%%%%%

\section{Method}
\label{sec:method}

% Summary of simulations
We performed four suites of $N$-body galaxy simulations: (A) a simulation with a live dark-matter halo without Sagittarius, (B) a simulation with a static dark halo without Sagittarius, (C) a simulation with a live dark halo with Sagittarius, and (D) the same as Model~C but with a Sagittarius assigned an unrealistically large mass. Models A and B constitute a case-control study that isolates the disk response arising specifically from the bar's slow-down (Section~\ref{sec:isolated}). In Models C and D, we introduce Sagittarius at a late stage of Model~A to examine how the bar's resonance becomes perturbed (Section~\ref{sec:Sagittarius}).

\subsection{Initial conditions}
\label{sec:initial_conditions}

The initial conditions of our galaxy models are generated using the \textsc{Galactics} code\footnote{\url{https://gitlab.com/jdubinski-group/GalactICS}} as described in \cite{Kuijken1995Nearly,Widrow2005Equilibrium,Widrow2008Blueprints}. The latest version of the code generalizes the previous methods to build galaxy models with the superposition of multiple disk and spherical components with density profiles specified by the user.  Gravitational potentials and distribution functions (DFs) are derived for each component on a logarithmic radial grid improving the accuracy over a large radial dynamic range. Monte-Carlo sampling of the DFs creates a $N$-body representation in approximate equilibrium of a multi-component galaxy containing a stellar disk, bulge and dark matter halo. The option also occurs to treat the dark matter halo as a static background potential within $N$-body simulations.

\subsubsection{Milky Way galaxy}
\label{sec:model_MW}

\begin{figure}
  \begin{center}
    \includegraphics[width=8.5cm]{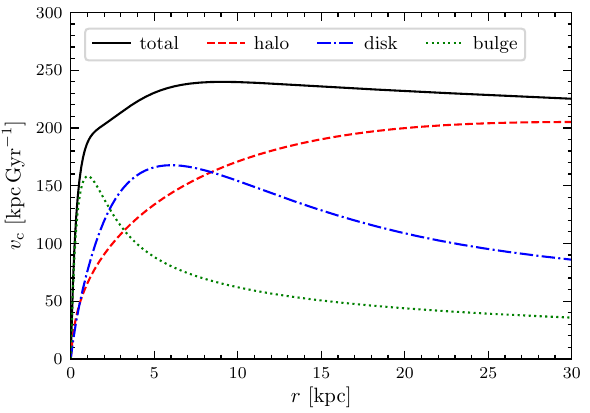}
    \caption{Rotation curve of the initial conditions of our Galaxy model.}
    \label{fig:rotation_curve}
  \end{center}
\end{figure}

Our Milky Way model is purely collisionless and consists of three components: a spherical NFW dark halo \citep{Navarro1997Universal}, an exponential stellar disk, and a spherical S\'{e}rsic stellar bulge \citep{sersic1968atlas,Prugniel1997fundamental}. Their target density profiles are
\begin{align}
  \rho_{\rm halo}(\vx) &= \frac{\rhoh}{(r/\rh)(1+r/\rh)^2}, \\
  \rho_{\rm disk}(\vx) &= \rhod \exp\left(- \frac{R}{\Rd}\right) \sech^2\left(\frac{z}{\zd}\right), \\
  \rho_{\rm bulge}(\vx) &= \rhob \left(\frac{r}{\rb}\right)^p \exp\left[ - b\left( \frac{r}{\rb} \right)^{\frac{1}{n}}\right],
  \label{eq:IC_density}
\end{align}
where $p = 1-0.6097/n - 0.05563/n^2$ and $b$ is adjusted such that $\rb$ is the projected half-mass radius. We adopt scale lengths of $\rh=15 \kpc$, $\Rd=2.8 \kpc$, $\zd=0.3 \kpc$ and $\rb=0.6 \kpc$, and set the S\'{e}rsic index to be $n=1$. The three components are smoothly truncated at $200 \kpc$, $20\kpc$ and $6\kpc$, respectively, using a logistic function. This makes the mass and radial extent of all models finite---this is especially important for the NFW profile which formally has infinite mass.
The density normalizations $\rho_{\rm h,d,b}$ are adjusted such that the total masses are $\Mh = 9 \times 10^{11} \Msun$, $\Md = 5 \times 10^{10} \Msun$ and $\Mh = 9 \times 10^{9} \Msun$, respectively. The resulting rotation curve is shown in Fig.\,\ref{fig:rotation_curve}. A detailed description of the model setup can be found in \citet{Widrow2008Blueprints}.

The disk particles are sampled from an approximate three-integral DF that depends on the energy in planar motions, the vertical energy, and the $z$-angular momentum $L_z$ \citep{Kuijken1995Nearly}. The square of the radial velocity dispersion is modeled to decline exponentially with radius: $\sigma^2_R(R)=\sigma^2_{R0} \exp(-R/R_\sigma)$. We adopt $\sigma_{R0} = 100 \kpcGyr$ and set $R_\sigma = \Rd$ for simplicity. The bulge particles are sampled from an ergodic DF, i.e. $f_E=f(E)$, while the halo is given a net rotation by adding to $f_E$ an odd function of the $z$-angular momentum $f_{\rm odd}$. Introducing rotation to the halo is essential, as it significantly affects the strength of dynamical friction on the bar \citep[e.g.][]{Athanassoula1996Evolution,weinberg1985evolution,Saha2013Spinning,Long2014Secular,Collier2018SpinningHalo,Fujii2019DryGalaxy,Kataria2022Effects,Li2023Stellar}; see \cite{Chiba2024halospin} for a physical explanation. Following \cite{Chiba2024halospin}, we adopt the following function:
\begin{align}
  f_{\rm odd}(\Lz,L,E) = \Lambda \tanh \left( \frac{\chi\Lz}{L} \right) f_E(E),
  \label{eq:odd_function}
\end{align}
where $\Lambda \in [-1,1]$ describes the degree of halo spin and $\chi$ determines how steeply the DF varies with $\Lz/L$. Note that adding $f_{\rm odd}$ only affects the halo's net angular momentum and leaves the halo's density distribution unchanged. The latest version of \textsc{Galactics} allows one to parametrize the rotation of any spherical component with this method. Throughout this work, we adopt $\Lambda=0.5$ and $\chi=3$. The corresponding dimensionless total angular momentum \citep{Peebles1971Rotation} is $\lambda = J_{\rm h}|E_{\rm h}^{1/2}|/(G\Mh^{5/2}) = 0.056$, where $J_{\rm h}$ and $E_{\rm h}$ are the total angular momentum and energy of the halo. For comparison, halos in the IllustrisTNG cosmological simulation have a median spin of $\lambda = 0.038$ \citep{Zjupa2017AngularMomentumIllustris}, and \cite{Obreja2022first} estimated the spin of the Milky Way's halo to be $\lambda = 0.061$ based on the correlation between the angular momentum of the dark and stellar components found in cosmological simulations.

\begin{table}
  \centering
  \caption{Initial parameters of the Sagittarius dwarf galaxy in Models~C and~D. 
  Both the stellar and dark-matter components follow NFW profiles.}
  \label{tab:Sgr_models}
  \begin{tabular}{lcccc}
    \hline\hline
    Model & Component & $r_{\rm s}$ [kpc] & $r_{\rm t}$ [kpc] & $M$ [${\rm M}_\odot$]\\
    \hline
    \noalign{\vskip 1pt}
    Model C 
    & Stars & 1 & 10 & $8.6\times10^{8}$\\
    & Dark matter & 8 & 20 & $2.5\times10^{10}$ \\
    \hline
    \noalign{\vskip 1pt}
    Model D 
    & Stars & 1 & 10 & $8.5\times10^{9}$ \\
    & Dark matter & 8 & 20 & $2.1\times10^{11}$ \\
    \hline
  \end{tabular}
\end{table}

\subsubsection{Sagittarius dwarf galaxy}
\label{sec:model_Sgr}

We used \textsc{Galactics} to generate a model dwarf galaxy as a superposition of two spherical NFW models, representing the stellar and dark matter components. The initial scale radius, truncation radius, and total mass are summarized in Table\,\ref{tab:Sgr_models}. Model~D adopts an order of magnitude larger mass than Model~C. We introduce Sagittarius at time $t=7 \Gyr$, well after the bar has formed. The initial phase-space coordinates of Sagittarius are determined by first integrating its orbit backward in time for $\Delta t=3$ Gyr from its observed present-day position and velocity as given by \cite{Vasiliev2020Sagittarius}. This backward integration is performed in a frozen axisymmetric potential at $t=7 \Gyr$, applying the Chandrasekhar's dynamical-friction formula \citep{Chandrasekhar1943DynamicalFriction} with a Coulomb logarithm of $\ln \Lambda=2$ and with a constant mass-loss rate of $(M_0 - M_1) /\Delta t$, where $M_0$ and $M_1$ are the initial and present-day mass, respectively (see \citealt{Jiang2000SagittariusDwarf} for a more elaborate treatment). Because the actual orbit of a tidally disrupting satellite deviates from that predicted by the Chandrasekhar's formula \citep{Fujii2006Dynamical}, we run a couple of low-resolution simulations to fine-tune the initial parameters until satisfactory convergence is achieved. In both models, Sagittarius reaches approximately its present-day position after its third pericentric passage (Fig.\,\ref{fig:Sgr_trajectory}). The final total mass enclosed within $5\kpc$ is $9\times10^{8}\Msun$ in Model~C and $2\times10^{9}\Msun$ in Model~D, both substantially larger than the observationally inferred value of $4\times10^{8}\Msun$ \citep{Vasiliev2020Sagittarius}. We intentionally adopt these large masses in order to test the robustness of the bar's resonant structure. Further details are provided in Section~\ref{sec:Sagittarius}.

\subsection{$N$-body simulation}
\label{sec:nbody_simulation}

We integrate the particles using the \textsc{Gadget-4} code, last described in \cite{Springel2021GADGET4}. Each model consists of $N = 10^7$ particles in both the dark-matter halo and the stellar disk, and $N = 2 \times 10^6$ particles in the bulge. The particle masses of the halo, disk, and bulge components are $9 \times 10^4 \Msun$, $5 \times 10^3 \Msun$, and $4.5 \times 10^3 \Msun$, respectively. The gravitational softening lengths are set to $20 \pc$ for stellar particles and $100 \pc$ for dark-matter particles. We adopt a hierarchical time integration with a maximum time step of $50 \Myr$.

\subsection{Angle-action coordinates}
\label{sec:angle_action_coordinates}

We explore the phase-space structure of the bar's resonance using angle-action variables $(\vtheta,\vJ)$, which constitute a set of canonical coordinates \citep[e.g.][]{binney2008galactic}. For axisymmetric galaxies, a common choice of actions is $\vJ=(\Jphi,\Jr,\Jz)$, where $\Jphi$ is the $z$-component of the angular momentum $\Lz$, $\Jr$ measures the radial excursion (eccentricity) of the orbit, and $\Jz$ describes the extent of vertical motion about the disk mid-plane. The conjugate angles $\vtheta=(\thetaphi,\thetar,\thetaz)$ specify the phase of the azimuthal, radial, and vertical motion, respectively, and their rates of change define the orbital frequencies $\vOmega=\dot{\vtheta}=(\Omegaphi,\Omegar,\Omegaz)$. We transform positions and velocities to angle-action coordinates using the St\"ackel fudge \citep{Binney2012Stackel} and perform the inverse transform using the torus mapper \citep{McGill1990Torus,Kaasalainen1994Torusconstruction}, both as implemented in \textsc{Agama} \citep{Vasiliev2019AGAMA}.

The phase-space dynamics near a resonance is best viewed by making an additional canonical transformation to the slow-fast angle-action coordinates defined for each resonance $\vN=(\Nphi,\Nr,\Nz)$ \citep[e.g.][]{LyndenBell1973Topics,Tremaine1984Dynamical,Binney2016Managing,binney2017orbital,monari2017distribution,Chiba2022Oscillating}. Again, the definitions are not unique, but here we choose 
\begin{align}
  &\thetas = \vN \cdot \vtheta - \Nphi \phib, ~ \thetafo = \thetar, ~ \thetaft = \thetaz, \\
  &\Js = \frac{\Jphi}{\Nphi}, ~ \Jfo = \Jr - \frac{\Nr}{\Nphi}\Jphi, ~ \Jft = \Jz - \frac{\Nz}{\Nphi}\Jphi,
  \label{eq:slowfastAA}
\end{align}
where $(\thetas,\Js)$ are the `slow' angle-action variables, and $(\vthetaf,\vJf) = (\thetafo,\thetaft,\Jfo,\Jft)$ are the `fast' angle-action variables. Here $\phib$ denotes the azimuthal coordinate of the bar's major axis. The slow variables describe the nonlinear secular evolution that occurs near the resonance, as implied by the resonance condition $\dot{\theta}_{\rm s} = \vN \cdot \vOmega - \Nphi \dot{\varphi}_{\rm b}=0$. The motion in this phase space is similar to that of a pendulum, where the phase space is split by the separatrix into two distinct regimes: trapped orbits lie within the separatrix and `librate' (their slow angle oscillates about the resonance), while untrapped orbits lie outside the separatrix and `circulate' (their slow angle increases monotonically); see e.g. Fig.\,3 of \citealt{Chiba2022Oscillating}. During this slow evolution, the fast angles $\vthetaf$ evolve rapidly and the fast actions $\vJf$ are conserved on average. In this paper, we focus on the bar's corotation resonance $\vN=(2,0,0)$, for which the slow angle-action is essentially the azimuthal angle-action in the bar's rotation frame $(\thetas,\Js)=(\Nphi\thetaphi',\Jphi/\Nphi)$, where $\thetaphi' \equiv \thetaphi-\phib$ is the azimuthal phase with respect to the bar angle. The fast angle-actions are simply the radial and vertical angle-actions: $(\vthetaf,\vJf) = (\thetar,\thetaz,\Jr,\Jz)$.

In idealized models of barred galaxies with $\pi$-period (two-fold) rotational symmetry, i.e. $\Phi(\varphi)=\Phi(\varphi+\pi)$, the dynamics on one side of the bar is identical to that on the opposite side, making it natural to use $\thetas=2\thetaphi' \in [0,2\pi]$, which only explores $\thetaphi' \in [0,\pi]$. In $N$-body simulations, however, the dynamics is not perfectly rotationally symmetric (the potential contains odd azimuthal Fourier components $m=1,3,5 \ldots$), owing to the finite number of particles which are initially sampled randomly in phase. We therefore work with $(\thetaphi',\Jphi)$ instead of $(\thetas,\Js)$. Henceforth, $(\vtheta,\vJ)$ denotes $(\thetaphi',\vthetaf,\Jphi,\vJf)$.

\section{Results}
\label{sec:results}

\subsection{Tree-ring structure in isolated barred galaxy}
\label{sec:isolated}

We begin by investigating the phase-space distribution of stars in an isolated galaxy. To assess the impact of the bar's spin-down, we compare simulations with a live halo to those with a static halo (i.e. a fixed halo potential), keeping all other aspects of the model identical. In the latter case, the halo does not exert dynamical friction on the bar and therefore the bar retains a constant pattern speed. The time evolution of the bar's properties is provided in Appendix \ref{sec:bar}.

\begin{figure*}
  \begin{center}
    \includegraphics[width=18.0cm]{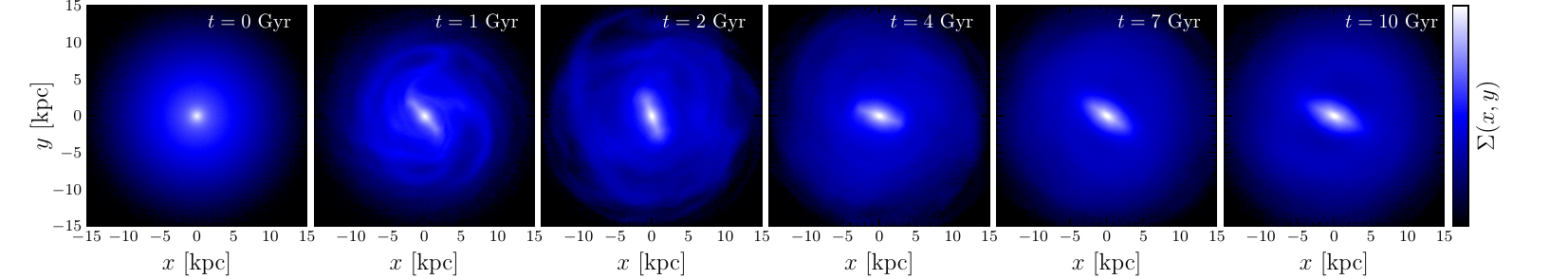}
    \includegraphics[width=18.0cm]{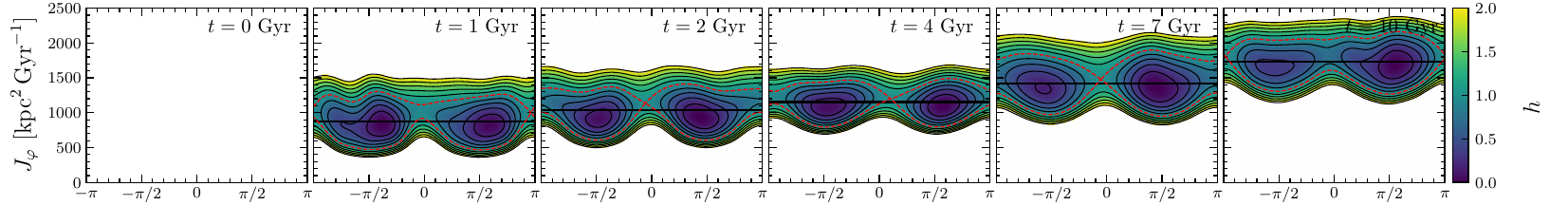}
    \includegraphics[width=18.0cm]{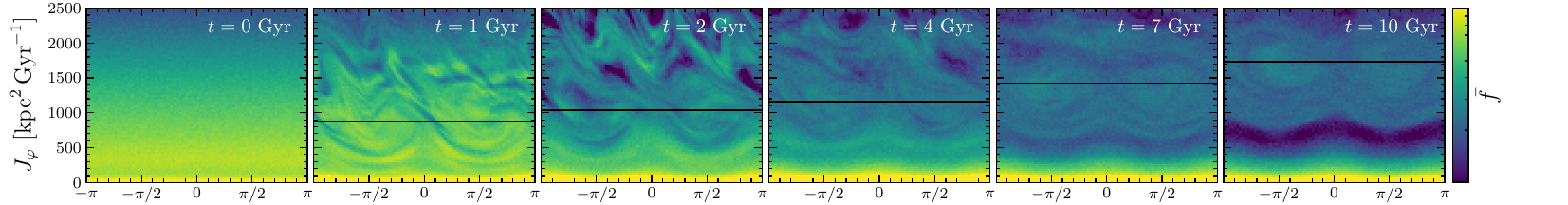}
    \includegraphics[width=18.0cm]{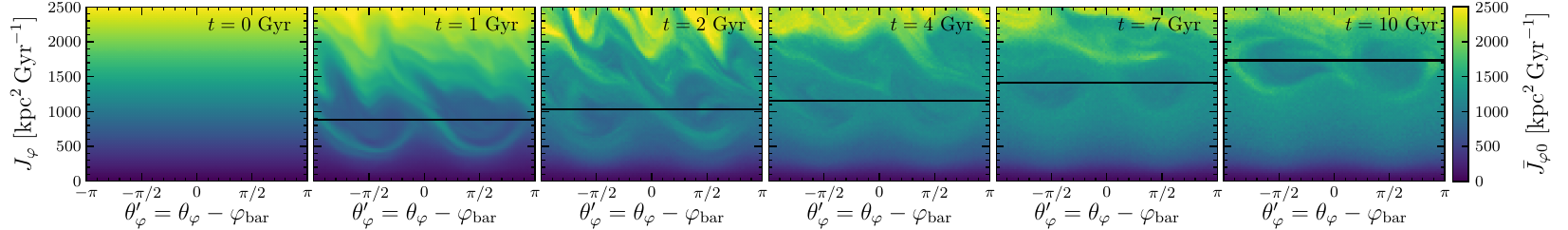}
    \caption{Snapshots of an isolated $N$-body galactic disk in a live dark halo (Model~A). Top row: Stellar surface density. Second row: Dimensionless fast-angle-averaged Hamiltonian (equation \ref{eq:h}) in the azimuthal angle-action space at $\vJf=(10,10) \kpckpcGyr$. Third and fourth rows: Density and initial angular momentum of stars averaged over $\vthetaf \in [0,2\pi]^2$ and $\vJf \in [0,100]^2 \kpckpcGyr$. As the bar spins down, the bar's corotation resonance (black lines) sweeps toward large $\Jphi$ and the trapped stars get dragged along with it.}
    \label{fig:xy_f_Jphi0_H}
  \end{center}
\end{figure*}

\begin{figure*}
  \begin{center}
    \includegraphics[width=18.0cm]{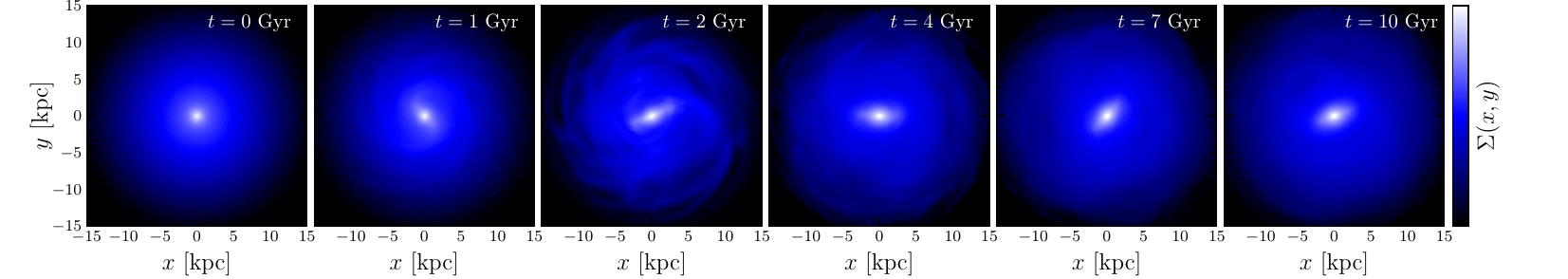}
    \includegraphics[width=18.0cm]{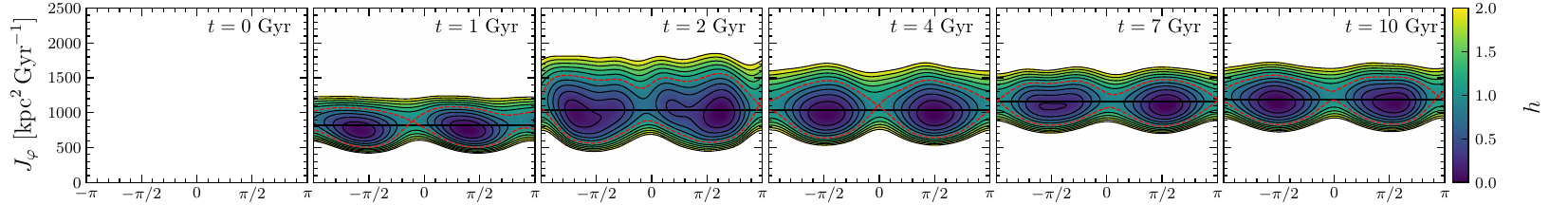}
    \includegraphics[width=18.0cm]{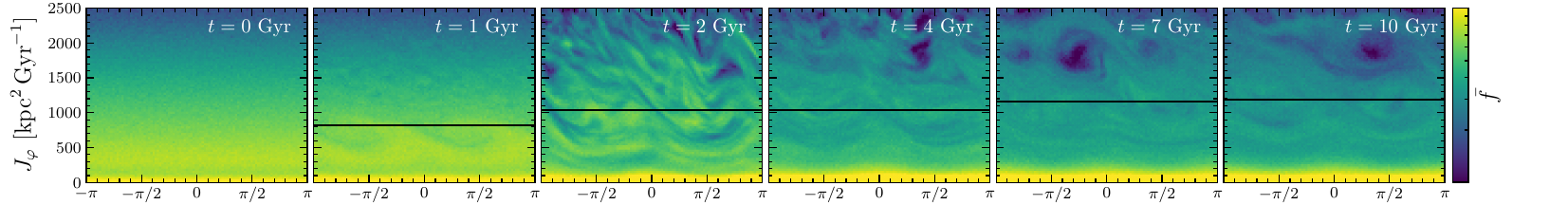}
    \includegraphics[width=18.0cm]{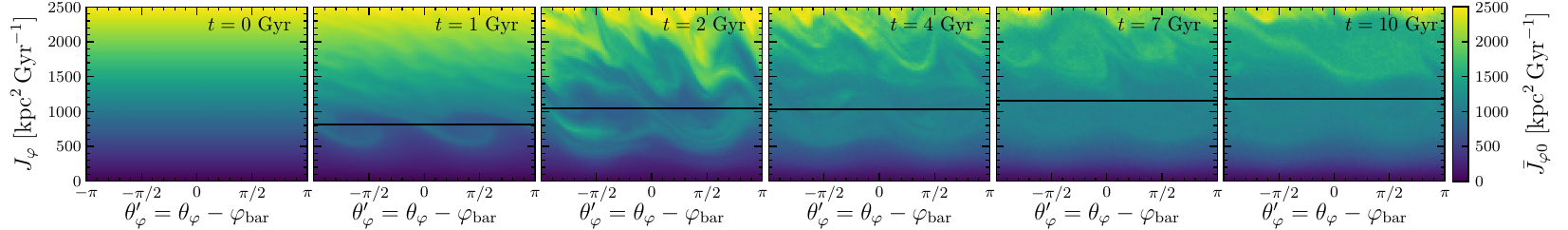}
    \caption{As in Fig.\,\ref{fig:xy_f_Jphi0_H}, but with a static dark halo (Model~B). Since the bar maintains a roughly constant pattern speed after $t\simeq 2 \Gyr$ (Fig.~\ref{fig:bar_properties}), the bar's resonance is kept fixed and therefore does not develop any distinct contrast with the surrounding phase space.}
    \label{fig:xy_f_Jphi0_H_static}
  \end{center}
\end{figure*}

Figure\,\ref{fig:xy_f_Jphi0_H} presents snapshots of the $N$-body simulation with a live dark halo at times $t=0,1,2,4,7$ and $10 \Gyr$ as indicated at the upper right corner of each panel. The top row of Fig.\,\ref{fig:xy_f_Jphi0_H} shows the integrated surface density of stars on a logarithmic scale. The bar forms within a Gyr and gradually grows in length and strength (cf. Fig.~\ref{fig:bar_properties}). 

The following rows present the simulation in the azimuthal (slow) angle-action space $(\thetaphi',\Jphi)$. To guide the reader, we first show in the second row of Fig.\,\ref{fig:xy_f_Jphi0_H} the Hamiltonian averaged over the fast angles
\begin{align}
  \bH(\thetaphi',\Jphi,\vJf) \equiv& \frac{1}{(2\pi)^2} \int \drm^2 \vthetaf \,H(\vtheta,\vJ) \nn \\
  =& H_0(\vJ) - \Omegap \Jphi + \frac{1}{(2\pi)^2} \int \drm^2 \vthetaf \,\Phi_1(\vtheta,\vJ),
  \label{eq:aveH}
\end{align}
where $H_0=\vvel^2/2+\Phi_0(\vx)$ is the Hamiltonian associated with the axisymmetric potential $\Phi_0$, $\Omegap$ is the bar's pattern speed, and $\Phi_1\equiv \Phi-\Phi_0$ denotes the non-axisymmetric component of the potential. The term $-\Omegap \Jphi$ appears in the Hamiltonian because we are in the bar's rotating frame. The computation of $\bH$ requires the transformation $(\vtheta,\vJ)\rightarrow(\vx,\vvel)$, which is performed using the torus mapper \citep{McGill1990Torus,Kaasalainen1994Torusconstruction}. The potential is evaluated using the \textsc{AGAMA} library, with a multipole expansion for the halo and a combination of Fourier expansion in $\varphi$ and a grid-based spline interpolation in $R$ and $z$ for the disk. The second row of Fig.\,\ref{fig:xy_f_Jphi0_H} shows $\bH$ at $\vJf=(10,10) \kpckpcGyr$, measured with respect to its value at the resonance center and normalized by its value at the separatrix:
\begin{align}
  h(\thetaphi',\Jphi,\vJf) = \frac{\bar{H}(\thetaphi',\Jphi,\vJf)-\bHres}{\bHsep-\bHres}, 
  \label{eq:h}
\end{align}
where $h=0$ corresponds to the resonance center and $h=1$ corresponds to the separatrix. We define $\bHres$ as the maximum value of $\bH$, and $\bHsep$ as the value at the saddle point of $\bH$. Since there are two saddle points that may take slightly different values of $\bH$, we adopt the smaller of the two as $\bHsep$. The figures show contours of $h$ from 0 to 2 at intervals of $\Delta h=0.2$, where the separatrix $h=1$ is marked with red dashed curves. The black horizontal lines mark the position of the bar's corotation resonance $\Omega=\Omegap$. We do not plot the Hamiltonian at $t=0$ since the bar has not yet developed and hence the pattern speed is ill-defined.

The fast-angle-averaged Hamiltonian shows the expected resonant phase-space structure: stars near the resonance librate about the resonance center in an anticlockwise manner, while those farther away circulate freely, with opposite directions above and below the resonance. The resonance gradually moves up as the bar spins down. Because this bar slowdown happens at a timescale longer than the typical period of libration, the majority of trapped stars are expected to adiabatically conserve the phase-space area enclosed by these contours (i.e. the libration action) and therefore get dragged toward larger angular momentum \citep{Chiba2020ResonanceSweeping}.

The third row of Fig.\,\ref{fig:xy_f_Jphi0_H} shows the distribution of stars in this azimuthal angle-action space, obtained by marginalizing the DF $f$ over the fast angles $\vthetaf=(\thetar,\thetaz)$ and the fast actions $\vJf=(\Jr,\Jz)$ within the domain $\Gamma = [0,100]\times[0,100]\kpckpcGyr$:
\begin{align}
  \bar{f}(\thetaphi',\Jphi) = \int \drm^2 \vthetaf \int_\Gamma \drm^2 \vJf \,f(\vtheta,\vJ).
  \label{eq:f_average}
\end{align}
Here we require the transformation $(\vx,\vvel)\rightarrow(\vtheta,\vJ)$, which is performed using the St\"ackel fudge \citep{Binney2012Stackel}. The distribution evolves approximately along the contours of $h$ and develops features that closely align with them (animations are available online\footnote{\label{fn:animation}\url{https://rimpeichiba.github.io/movies/2026_Nbody/}}). The phase-space density of the trapped region is clearly higher than that of the surrounding phase space, indicating that stars trapped in the inner dense disk have been transported to the outer disk.\footnote{Note that the significant depletion of stars at $\Jphi \sim 600 \kpckpcGyr$ is mostly due to the inner Lindblad resonance which transports stars toward large $\Jr$ outside of our sampling domain $\Gamma$.}

This is clarified in the fourth row of Fig.\,\ref{fig:xy_f_Jphi0_H}, which shows the average initial angular momentum of stars:
\begin{align}
  \bar{J}_{\varphi 0}(\thetaphi',\Jphi) = \bar{f}^{-1} \int \drm^2 \vthetaf \int_\Gamma \drm^2 \vJf \,f(\vtheta,\vJ) \,J_{\varphi 0}(\vtheta,\vJ).
  \label{eq:Jphi0_average}
\end{align}
The initial angular momentum is significantly lower at the resonance, confirming the systematic migration of trapped stars due to the bar's spin-down. Such extreme stellar migration induced by a sweeping resonance has already been indicated in earlier $N$-body simulations \citep{Dubinski2009Anatomy,Halle2018Radial,Khoperskov2019Escapees}.

Figure\,\ref{fig:xy_f_Jphi0_H_static} shows the corresponding results for the simulation with a static halo. After $t \simeq 2 \Gyr$, the position of the bar's resonance remains almost fixed because the bar ceases to spin down (cf. Fig.~\ref{fig:bar_properties}). Due to the absence of resonance sweeping, no distinct overdensity or angular momentum contrast develops at the resonance.

The key prediction of \cite{Chiba2021TreeRing} was the emergence of a tree-ring structure inside the resonance of a slowing bar. If the phase-space volume of trapping increases as it sweeps radially outward (as naturally happens when the bar grows while slowing down), it will sequentially capture stars at the expanding surface. Since the trapped stars adiabatically conserve the libration action,
\begin{align}
  \Jl = \frac{1}{2\pi} \oint \drm \thetaphi' \Jphi,
  \label{eq:Jl}
\end{align}
which measures the phase-space area enclosed by the trapped orbits, they approximately preserve their relative ordering within the resonance. This leads to a characteristic structure in which the libration action correlates with the stars' capture radii and thus also with their birth radii (or their initial angular momenta).

\begin{figure}
  \begin{center}
    \includegraphics[width=8.5cm]{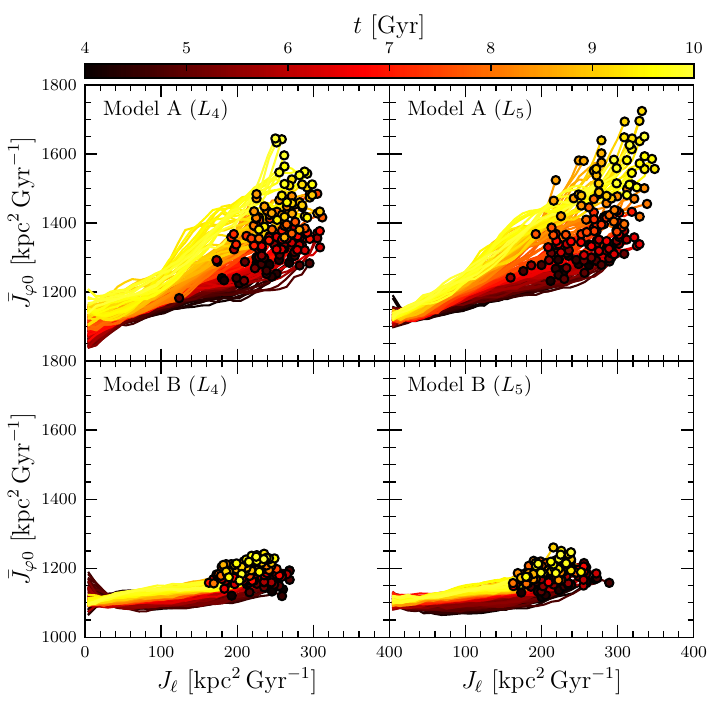}
    \caption{Mean initial angular momentum $\bar{J}_{\varphi 0}$ of stars trapped in the bar's corotation resonance as a function of their libration action $J_{\ell}$, color-coded by time. The top panels show results for the live-halo model (Model A), in which the bar slows, while the bottom panels show those for the static-halo model (Model B), in which the bar rotates steadily. The two columns present results for the two distinct resonant islands centered on the Lagrange points $L_4$ and $L_5$. Open circles mark the endpoint (separatrix) at each epoch. The resonance in Model A exhibits a positive gradient in $\bar{J}_{\varphi 0}$ which becomes steeper over time, since the resonance sequentially captures new stars at larger angular momentum as it sweeps and expands in volume. In contrast, the resonance in Model B shows an almost constant flat distribution.}
    \label{fig:Jp0_Jl}
  \end{center}
\end{figure}

We demonstrate this in Fig.\,\ref{fig:Jp0_Jl}, which shows the correlation between the libration action $\Jl$ and the mean initial angular momentum $\bar{J}_{\varphi 0}$, color-coded by time. Details of the numerical method to compute $\Jl$ are given in Appendix~\ref{sec:computation_of_Jl}. Since the bar's corotation consists of two resonant islands centered on the stable Lagrange points, commonly denoted as $L_4$ and $L_5$, we present the results for each island separately, with $L_4$ shown on the left and $L_5$ on the right.

The top panels of Fig.\,\ref{fig:Jp0_Jl} show the results for the live-halo model (Model~A). We see the expected tree-ring structure: $\bar{J}_{\varphi 0}$ increases substantially with $\Jl$. The open circles mark the endpoints of each curve, i.e. the separatrix. Their evolution toward the upper right indicates that the resonance grows in volume on average\footnote{In our model, the resonance exhibits limited secular growth, because the simulation starts with a fully grown disk, in which the bar develops rapidly and reaches near-maximum strength at early times (cf. Fig.~\ref{fig:bar_properties}). In more realistic models with a gradually growing disk, the bar and its resonance are expected to grow more steadily \citep[e.g.][]{Aumer2016Age,Dehnen2023Measuring}.} while simultaneously sweeping toward larger $\Jphi$. In an idealized model without stellar diffusion, the internal structure would be preserved, and the curves would extend along the trajectory of these endpoints.\footnote{This is subject to the condition that the resonance moves adiabatically. If the resonance sweeps fast, the effective resonant volume, within which stars remain trapped, shrinks \citep{Chiba2023GeneralFastSlowRegime}.} In practice, however, trapped stars diffuse both within the resonance and across the separatrix. Consequently, newly captured stars with high $J_{\varphi 0}$ can gradually migrate into the core of the resonance, causing the inner part of the curves to rise. This smooths and straightens the curves, although it does not erase the gradient entirely.

The bottom panels of Fig.\,\ref{fig:Jp0_Jl} show the corresponding results for the static-halo model (Model~B). The structures exhibit markedly different behavior: $\bar{J}_{\varphi 0}$ depends only weakly on $\Jl$ since the bar's resonance neither moves nor grows. We nevertheless find a slight positive correlation. The reason for this is rather technical and off topic, so we defer a detail explanation to Appendix~\ref{sec:asymmetry_Hamiltonian}. In brief, the positive correlation emerges because the Hamiltonian is slightly asymmetric about the resonance (Fig.\,\ref{fig:xy_f_Jphi0_H_static}). This asymmetry causes stars with higher angular momentum to be preferentially sampled at large $\Jl$, resulting in a weak positive correlation.

\begin{figure}
  \begin{center}
    \includegraphics[width=8.5cm]{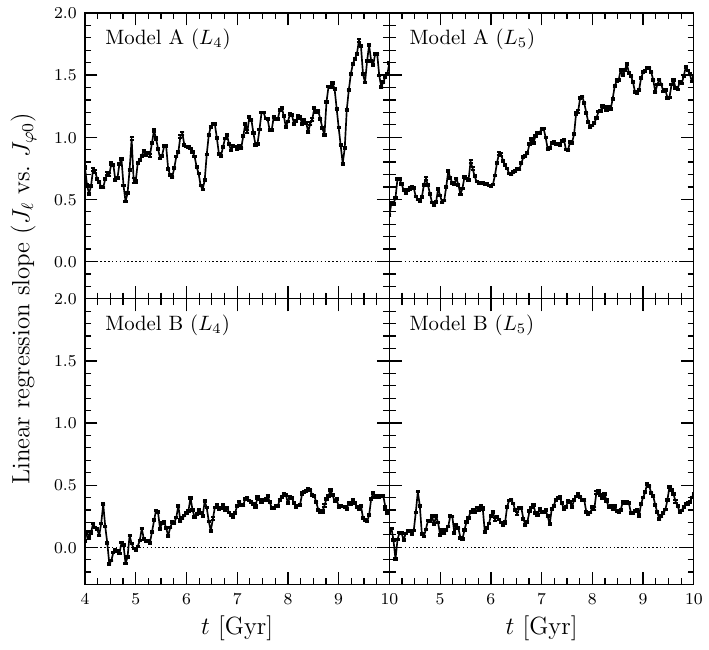}
    \caption{Linear regression slope between $J_{\ell}$ and $J_{\varphi 0}$ as a function of time. The slope increases significantly in the live-halo model (Model A), whereas it remains nearly constant in the static-halo model (Model B).}
    \label{fig:dJp0dJl_ModelAB}
  \end{center}
\end{figure}

To further clarify the time dependence of the tree-ring structure, we present in Fig.\,\ref{fig:dJp0dJl_ModelAB} the slope of $J_{\varphi 0}(\Jl)$ as a function of time. We obtain the slope from a mass-weighted linear regression of trapped stars in the $(\Jl,J_{\varphi 0})$ plane. Consistent with Fig.\,\ref{fig:Jp0_Jl}, the regression slope in Model~A (live halo) increases markedly with time. Model~B (static halo) also shows a modest increase in the slope up until $t\sim7 \Gyr$, although the evolution is distinctly smaller than in Model~A.

This positive correlation between $\Jl$ and $J_{\varphi 0}$ is precisely that predicted by \cite{Chiba2021TreeRing} in a simplified two-dimensional test-particle model and identified in the Milky Way using stellar metallicity as a proxy for $J_{\varphi 0}$. The identification relied on the assumption that the Hercules stellar stream in the Solar neighborhood is shaped by the bar's corotation resonance, an interpretation which has gained increasing support in recent years \citep{perez2017revisiting,Monari2019signatures,DOnghia2020Trojans,Binney2020Trapped,Chiba2020ResonanceSweeping,Kawata2021GalacticBarHotStar,Lucchini2024Milky,Dillamore2024Radial,Dillamore2025Dynamical,Khalil2025nonaxisymmetric,Li2025originHerculesII} and is consistent with the recent downward revisions on the measurements of the bar's current pattern speed \citep[e.g.][]{Clarke2021ViracGaia,Leung2023measurement,Zhang2024Kinematics}. Furthermore, \cite{Chiba2021TreeRing} argued that the tree-ring structure should emerge only when the correct pattern speed is assumed, and used this requirement to constrain the present-day pattern speed, obtaining values consistent with other independent measurements. In Appendix~\ref{sec:infer_wp}, we validate this method with our simulations and demonstrate that the true pattern speed can indeed be recovered by demanding a tree-ring structure inside the resonance.

\subsection{Tree-ring structure perturbed by the Sgr dwarf galaxy}
\label{sec:Sagittarius}

Having demonstrated that the resonance of a slowing bar exhibits a tree-ring structure in a self-consistent $N$-body simulation, we now investigate whether this structure persists in the presence of perturbations from the Sagittarius (Sgr) dwarf galaxy.

\begin{figure}
  \begin{center}
    \includegraphics[width=8.5cm]{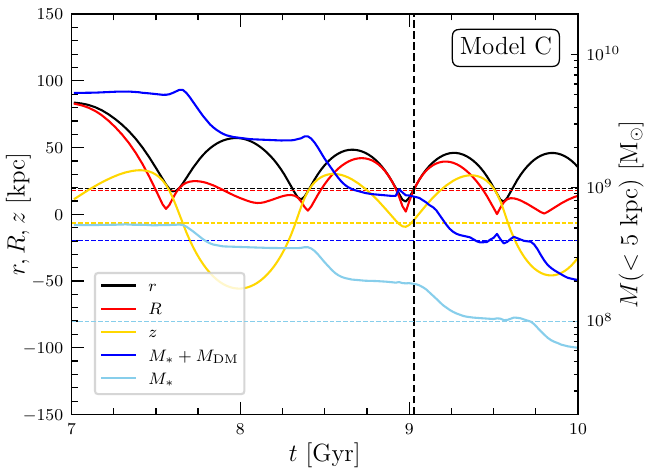}
    \includegraphics[width=8.5cm]{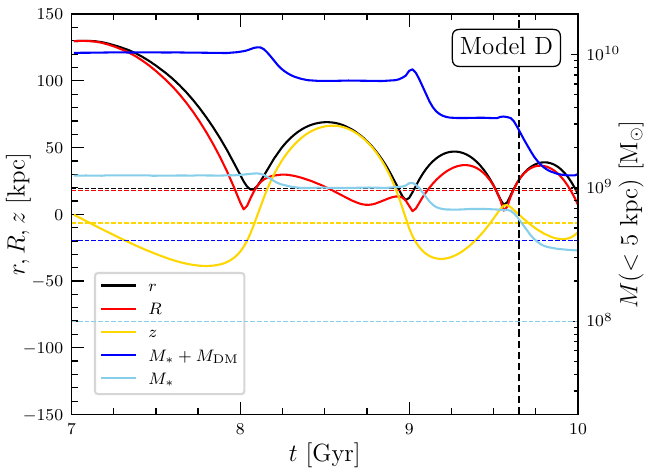}
    \caption{Time evolution of the position of the Sagittarius dwarf galaxy (left axis) and its mass enclosed within $5 \kpc$ from its density peak (right axis). Top panel shows the fiducial model (Model~C), while the bottom panel shows the high-mass Sagittarius model (Model~D). The horizontal dashed lines indicate the present-day values \citep{Vasiliev2020Sagittarius}, and the vertical dashed lines indicate the time when Sagittarius is closest to its present-day position. We intentionally chose models with large present-day mass so as not to underestimate its impact on the bar's resonance.}
    \label{fig:Sgr_trajectory}
  \end{center}
\end{figure}

As described in Section \ref{sec:model_Sgr}, we run two simulations (Models C and D) in which Sagittarius, with different initial masses and phase-space coordinates, is introduced at $t=7\Gyr$ into the fiducial live-halo model (Model~A; Fig.\,\ref{fig:xy_f_Jphi0_H}). Their trajectories and mass enclosed within $5\kpc$ are shown in Fig.\,\ref{fig:Sgr_trajectory}. The horizontal dotted lines indicate the present-day values, while the vertical dashed line marks the time when Sagittarius is approximately at its present-day position, as measured by \cite{Vasiliev2020Sagittarius}. In both models, Sagittarius reaches its present-day position after its third pericentric passage. However, the present-day mass in both models is significantly larger than the observed value: by a factor of $\sim 2$ in Model~C, and a factor of $\sim 8$ in Model~D. Since our purpose is to investigate the robustness of the bar's resonant structure, this overestimate is conservative: if the resonance survives in our models, it should also survive under a much weaker perturbation from the real Sagittarius.

\begin{figure*}
  \begin{center}
    \includegraphics[width=18.0cm]{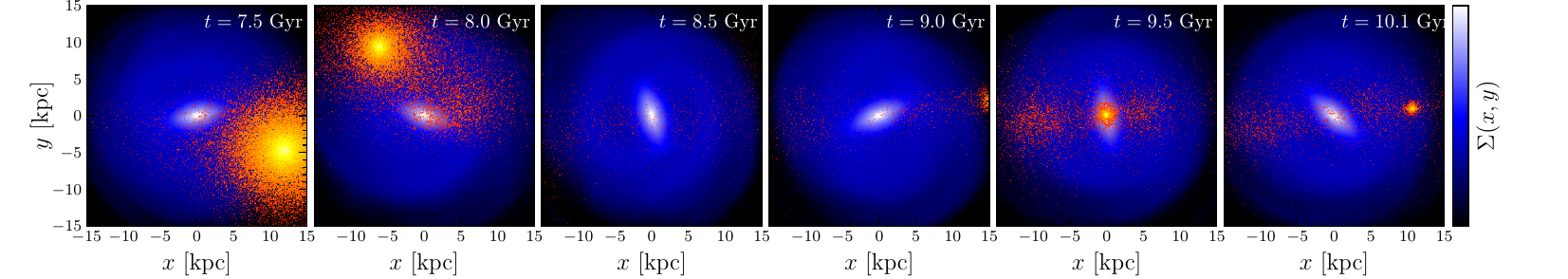}
    \includegraphics[width=18.0cm]{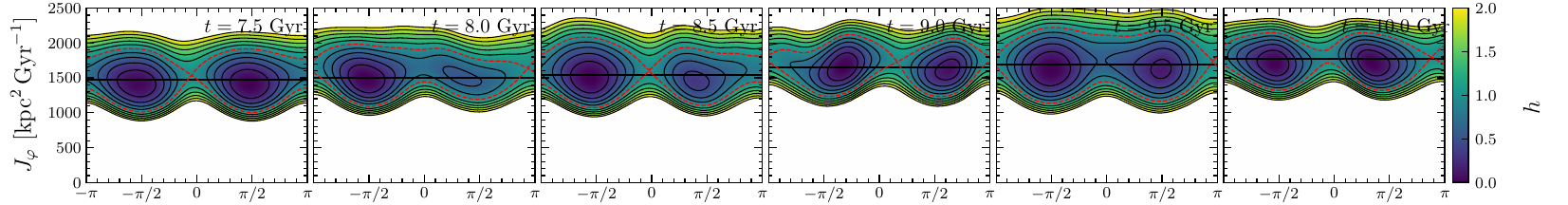}
    \includegraphics[width=18.0cm]{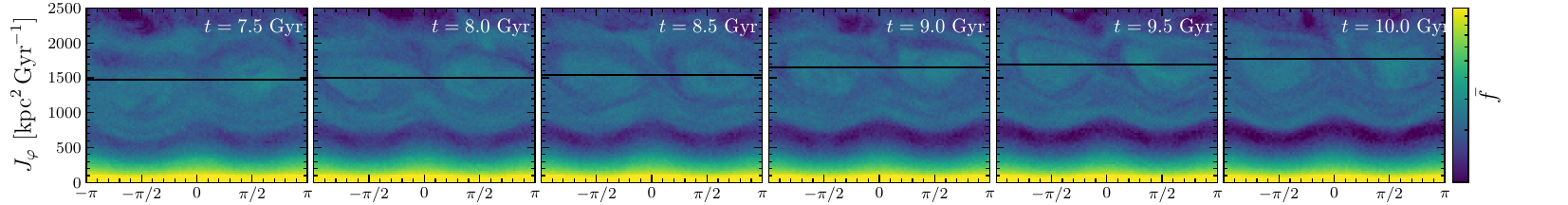}
    \includegraphics[width=18.0cm]{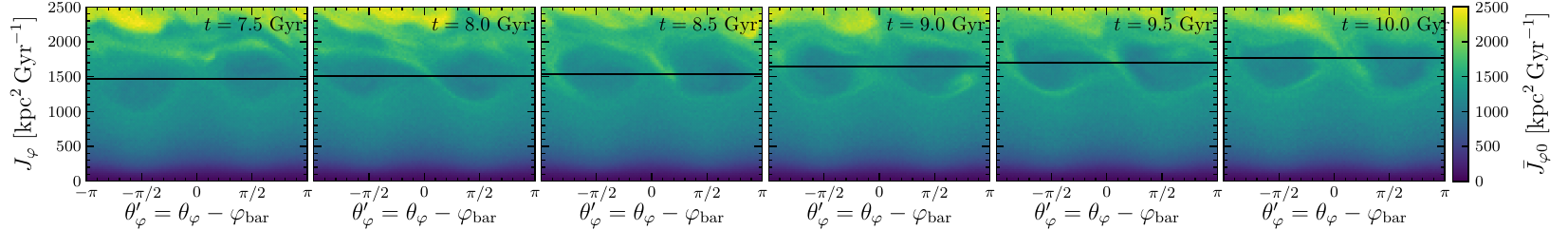}
    \caption{As in Fig.\,\ref{fig:xy_f_Jphi0_H}, but with a Sagittarius-dwarf galaxy introduced at $t=7 \Gyr$ (Model~C). In the top row, the surface density of Sagittarius and its dark halo is over plotted in red. The tidal force of Sagittarius induces a prominent two-armed spiral arm in the disk. Although the bar's resonant structure experiences significant large-scale fluctuations, it remains undestroyed. Animations are available online\hyperref[fn:animation]{\textsuperscript{\ref*{fn:animation}}}.}
    \label{fig:xy_f_Jphi0_H_Sgr}
  \end{center}
\end{figure*}

\begin{figure*}
  \begin{center}
    \includegraphics[width=18.0cm]{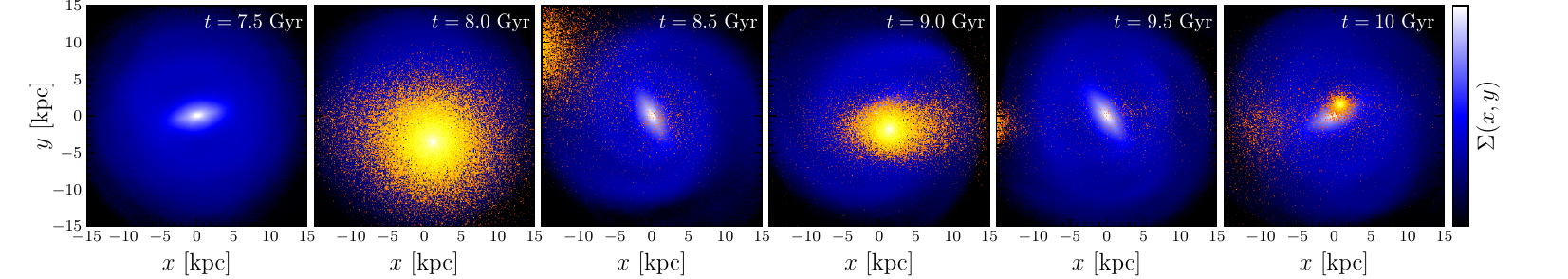}
    \includegraphics[width=18.0cm]{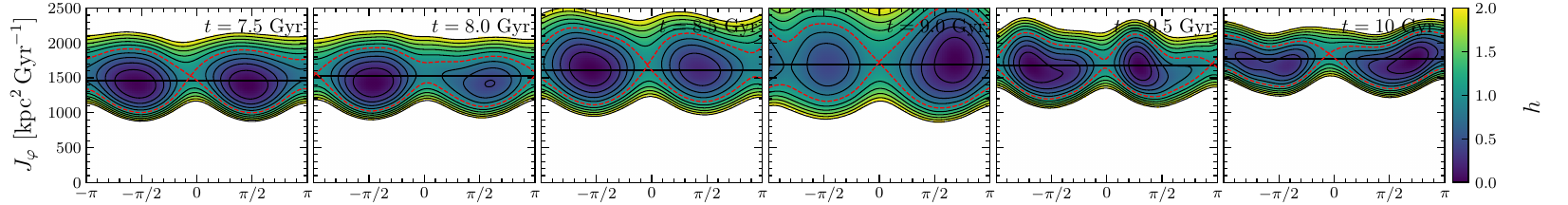}
    \includegraphics[width=18.0cm]{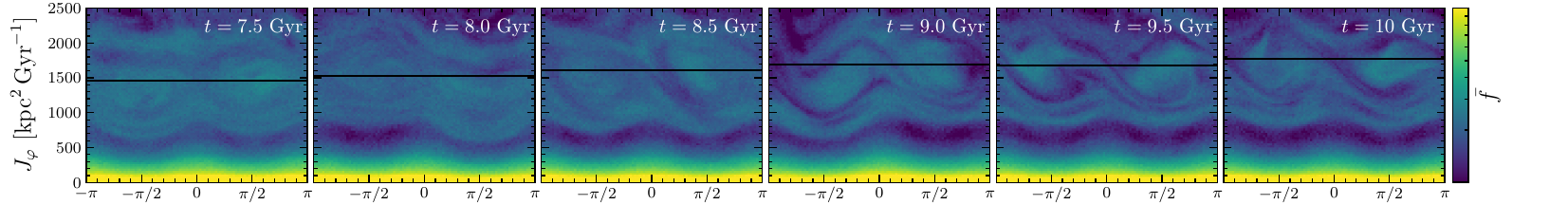}
    \includegraphics[width=18.0cm]{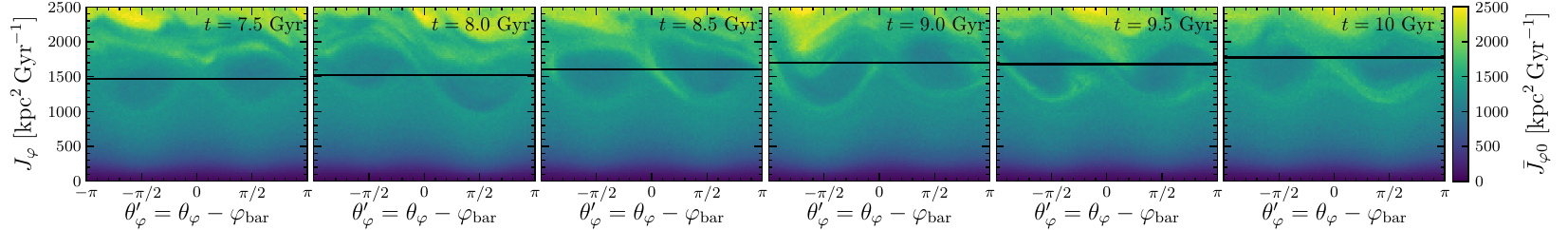}
    \caption{As in Fig.\,\ref{fig:xy_f_Jphi0_H_Sgr}, but with a more massive Sagittarius-dwarf galaxy (Model~D).}
    \label{fig:xy_f_Jphi0_H_Sgr_highmass}
  \end{center}
\end{figure*}

Figs.\,\ref{fig:xy_f_Jphi0_H_Sgr} and \ref{fig:xy_f_Jphi0_H_Sgr_highmass} show the results of Model~C and D, respectively. We present snapshots from $t=7.5$ to $10 \Gyr$ every $0.5 \Gyr$. In the top panels, the surface density of Sagittarius (sum of stellar and dark components) is over-plotted using a yellow-red color map. As Sagittarius crosses the disk, its tidal force induces a prominent two-armed spiral pattern in the disk. The subsequent plots in the slow angle-action plane show that the passage of Sagittarius generates large fluctuations in the bar's resonant structure (see animations\hyperref[fn:animation]{\textsuperscript{\ref*{fn:animation}}}). Nevertheless, its internal distribution is only weakly affected, and the characteristic tree-ring structure is preserved. Note that the transformation to and from angle-action coordinates is performed using the potential of the Milky Way alone and does not include the contribution from Sagittarius.

\begin{figure}
  \begin{center}
    \includegraphics[width=8.5cm]{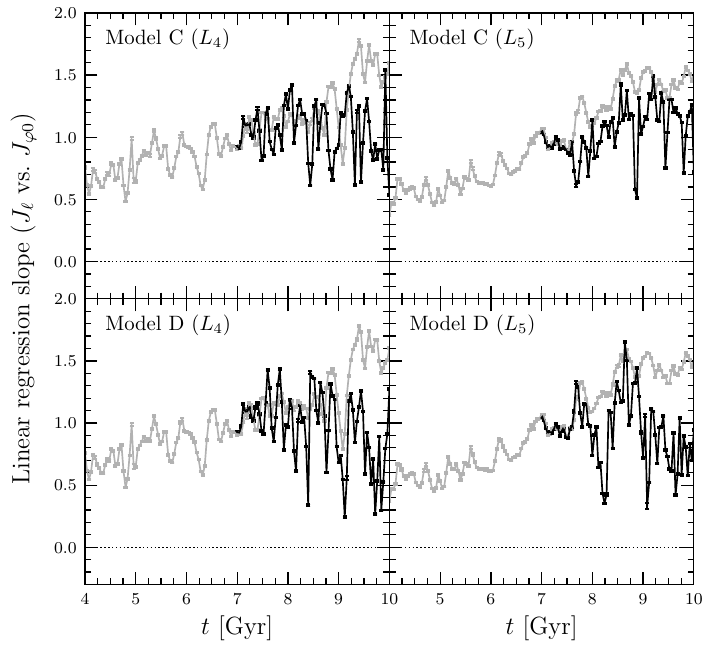}
    \caption{As in Fig.\,\ref{fig:dJp0dJl_ModelAB}, but comparing Models C and D with Model~A plotted in light gray. The regression slope is distorted and reduced by the impact of Sagittarius but nevertheless remains positive.}
    \label{fig:dJp0dJl_ModelCD}
  \end{center}
\end{figure}

Fig.\,\ref{fig:dJp0dJl_ModelCD} shows the regression slope of $J_{\varphi 0}(\Jl)$ inside the trapped phase-space, as in Fig.\,\ref{fig:dJp0dJl_ModelAB}. The correlation exhibits stochastic fluctuations, whose amplitude is larger in Model~D. We find that the slope is slightly reduced on average, indicating that stars have diffused in $\Jphi$. Nevertheless, a positive correlation between $\Jl$ and $J_{\varphi 0}$ clearly persists. We remind the reader that the present-day mass of Sagittarius is significantly overestimated in both models (by a factor of 8 in Model~D), so the expected impact in the real Galaxy is much smaller.

The remarkable robustness of the tree-ring structure can be attributed to at least three factors. First, as illustrated by \cite{Carr2022Migration}, the tidal perturbation from Sagittarius is strongest in the far outer disk and is relatively weak at the corotation radius ($R\sim 7 \kpc$). Second, the \textit{direct} impact of the tidal perturbation on the resonant structure is not very effective, as the spatial scale of the tide is larger than the bar's trapped region. As a result, the tide induces global fluctuations in the resonance without significantly affecting its internal structure. Third, its \textit{indirect} impact on the resonance through changes in the bar pattern speed is also limited, because there is no significant change to the averaged pattern speed over orbital timescales (Fig.\,\ref{fig:bar_properties}). Consequently, the resonance temporarily fluctuates but quickly returns to its original location before the resonant structure is disrupted.

\section{Discussion}
\label{sec:discussion}

This section discusses the limitations of our current model and highlights key physical ingredients that should be incorporated in future studies.

In this work, we modeled the Galaxy as a purely collisionless $N$-body system, neglecting the gaseous component, i.e. the interstellar medium (ISM). The ISM is known to be highly inhomogeneous at multiple scales as a result of several interrelated processes, including self-gravitational instability, supersonic turbulence, and stellar feedback \citep[e.g.][]{Meidt2023PHANGS,Beattie2025spectrum,Modak2025Characterizing}. Such inhomogeneities, e.g. giant molecular clouds (GMCs), can gravitationally scatter passing stars and are believed to play an important role in driving stellar diffusion in galactic disks \citep[e.g.][]{Spitzer1951Possible,Spitzer1953Possible,Jenkins1990Spiral,Aumer2016Age,Fujimoto2023Efficient}. This gas-driven stellar diffusion is particularly important in early gas-rich disks \citep[][]{Zhang2025Enhanced}. Including the ISM may thus enhance stellar diffusion, thereby weakening and potentially modifying the resonant structure \citep{Hamilton2023BarResonanceWithDiffusion,Chiba2025TwoArmed}. However, while GMCs play a crucial role in scattering stars vertically, their contribution to in-plane heating is minor \citep{Aumer2016QuiescentPhase,Aumer2016Age}, which likely limits their impact on the bar's resonance.

Furthermore, as in most previous simulations of a single isolated galaxy, our simulation begins with a fully grown, massive stellar disk and neglects ongoing star formation. Consequently, the disk is initially overly unstable, rapidly forming spiral arms and a bar. At later times, these very structures dynamically heat the disk, rendering it overly stable and suppressing further spiral activities. In real disk galaxies, continuous star formation introduces new stars on near-circular orbits, which cools the disk and counteracts the heating process, allowing spiral activities to persist for longer \citep[e.g.][]{Sellwood1984Spiral,Aumer2016QuiescentPhase}. Therefore, our models may have underestimated the impact of spiral perturbations on the bar's resonant structure at late times. 

The inclusion of gas and star formation is also required to model the chemical enrichment of the disk. In this study, we used the initial angular momentum of stars to trace their migration history. Observationally, however, this quantity can only be inferred indirectly from stellar chemical abundances, such as metallicity. Stellar metallicity provides a reliable proxy for initial angular momentum only for stars born at late times ($t_{\rm lookback} \lesssim 8 \Gyr$), after the radial profile of gas metallicity has largely stabilized \citep{Nordstrom2004GCS,Casagrande2011New,Schoenrich2017Understanding}. For those born during the early epoch of intense star formation and metal enrichment, the metallicity-birth radius relation would bear a strong age dependence. Hence, a proper interpretation of the observed tree-ring structure requires separating stars into coeval populations.

Finally, while this study focused on the birth radii of trapped stars, it would also be interesting to inspect their age distribution. Past simulations have shown that the formation of the bar triggers a burst of star formation in the bar region \citep[e.g.][]{Friedli1995Secular,Baba2022Age}. Since a fraction of these stars formed in the inner disk may subsequently migrate outward by surfing on the bar's corotation resonance, the age distribution of trapped stars may exhibit a distinct enhancement at the bar formation epoch \citep{Baba2025Influence}. In a follow-up study, we will examine the age-metallicity distribution of trapped stars using a fully chemo-dynamical galaxy simulation.

%%%%%%%%%%%%%%%%%%%%%%%%%%%%%%%%%%%%%%%%%%%%%%%%%%%%%%%%%%%%%%%%%%%%%%%%%%%%%%%%%%%%%%%%%%%%%%%%%%%%%%%%%%%%%%%%%%%%%%%%%%%%%%%%%%%%%%%%%%%%%%%%%%%%%%%%%%%%%%%%%%%%%%%%%%%%%%%%%%%%%%%%%%%%%%%%%%%%%%%%

\section{Summary}
\label{sec:Summary}

We investigated the phase-space evolution of stars trapped in the bar's corotation resonance using $N$-body simulations of a Milky Way-like galaxy. By inspecting the evolution in the azimuthal (slow) angle-action space, we showed that the spin-down of the bar gives rise to the predicted `tree-ring' structure, in which the initial angular momentum of trapped stars increases monotonically toward the separatrix of the resonance. This structure emerges because stars are captured sequentially from the inner to the outer disk as the resonance sweeps outward, and because trapped stars adiabatically conserve the libration action, which encodes the relative ordering within the resonance. We showed that this coherent ordering persists despite stellar diffusion within and across the separatrix, driven by stochastic fluctuations in the gravitational potential. We also confirmed that this structure is absent in simulations with a static dark halo, where the bar maintains an approximately constant pattern speed, demonstrating that the tree-ring structure is a unique dynamical signature of the deceleration of the bar.

We further demonstrated that this tree-ring structure remains remarkably robust in the presence of strong tidal perturbations from a Sagittarius-like dwarf galaxy. The passage of Sagittarius induces prominent tidal spiral arms and causes significant fluctuations in the bar's resonant structure. Despite the strong disturbance, however, the tree-ring structure remained undestroyed. We confirmed this robustness using two simulations with different initial mass and initial galactocentric radius for Sagittarius. In both models, the present-day mass of Sagittarius is significantly larger than the observed value, with one model overestimating it by nearly an order of magnitude, ensuring that the resulting tidal perturbations to the disk are not underestimated. We therefore conclude that the passage of Sagittarius in the real Galaxy is unlikely to have caused significant impact on the bar's resonant structure.

While our models self-consistently considered perturbations to the bar's resonance from spiral arms and the Sagittarius dwarf galaxy, they remain far from fully representing the messy environment of the real Galaxy. In particular, we have neglected the gaseous component, which constitutes an important source of small-scale fluctuations in the gravitational potential. We further ignored perturbations by other numerous satellites and dark subhalos \citep[see recent work by][]{Davies2026erasure}. Therefore, there still remains scope for additional perturbations that could further weaken or modify the bar's resonant structure.

Despite these limitations, however, the tree-ring structure has been observed in the Milky Way, implying that the Galactic bar has slowed in the past and that stellar diffusion has not been strong enough to erase this signature. The detailed morphology of this structure encodes valuable information about the dynamical nature of dark matter, the efficiency of stellar diffusion, and the history of chemical evolution in the Milky Way. This motivates future work to develop theoretical models that describe the chemo-dynamical evolution of the bar's resonant structure and to confront them with the rich observational data now available.

%% Please use the acknowledgment and contribution environments. This will 
%% be anonomyized when the "anonymous" style option is used. 
\begin{acknowledgments}

We are grateful to E. Vasiliev and T. Asano for their helpful advice on modeling the orbit of Sagittarius. RC is supported by the Japan Society for the Promotion of Science (JSPS) Research Fellowship, grant No. 25KJ0049.

Computations were performed in part on Cray XD2000 at the Center for Computational Astrophysics, National Astronomical Observatory of Japan and in part on the Niagara supercomputer at the SciNet HPC Consortium. SciNet is funded by Innovation, Science and Economic Development Canada; the Digital Research Alliance of Canada; the Ontario Research Fund: Research Excellence; and the University of Toronto.

\end{acknowledgments}

\begin{contribution}
%%This section gives authors the space to recognize author contributions. The text inside this environment is NOT counted towards the total word quanta. At a minimum, manuscripts are expected to include this text:

All authors contributed equally to the Terra Mater collaboration.

%% But authors are expected to provide more specific details, e.g. 
%%
%%SC was responsible for writing and submitting the manuscript.
%%WWM came up with the initial research concept and edited the manuscript.
%%OTS obtained the funding and edited the manuscript.
%%EBF provided the formal analysis and validation. He also edited the manuscript.
%%GEH Supervised the undergraduates, wrote the software and administers the project github and Zenodo repositories.
%%
%% Authors can use the Contributor Role Taxonomy (CRediT) at
%% https://credit.niso.org
%% for ideas on how write a good statement tailored to their needs.

\end{contribution}

%% To help institutions obtain information on the effectiveness of their 
%% telescopes the AAS Journals has created a group of keywords for telescope 
%% facilities.
%
%% Following the acknowledgments section, use the following syntax and the
%% \facility{} or \facilities{} macros to list the keywords of facilities used 
%% in the research for the paper.  Each keyword is check against the master 
%% list during copy editing.  Individual instruments can be provided in 
%% parentheses, after the keyword, but they are not verified.
\facilities{HST(STIS), Swift(XRT and UVOT), AAVSO, CTIO:1.3m, CTIO:1.5m, CXO}

%% Similar to \facility{}, there is the optional \software command to allow 
%% authors a place to specify which programs were used during the creation of 
%% the manuscript. Authors should list each code and include either a
%% citation or url to the code inside ()s when available.
%\software{astropy \citep{2013A&A...558A..33A,2018AJ....156..123A,2022ApJ...935..167A},  
%          Cloudy \citep{2013RMxAA..49..137F}, 
%          Source Extractor \citep{1996A&AS..117..393B}
%          }

%% Appendix material should be preceded with a single \appendix command.
%% There should be a \section command for each appendix. Mark appendix
%% subsections with the same markup you use in the main body of the paper.
%%
%% Each Appendix (indicated with \section) will be lettered A, B, C, etc.
%% The equation counter will reset when it encounters the \appendix
%% command and will number appendix equations (A1), (A2), etc. The
%% Figure and Table counter will not reset.

\appendix

\section{Analysis of the bar}
\label{sec:bar}
\begin{figure}
  \begin{center}  
    \includegraphics[width=8.5cm]{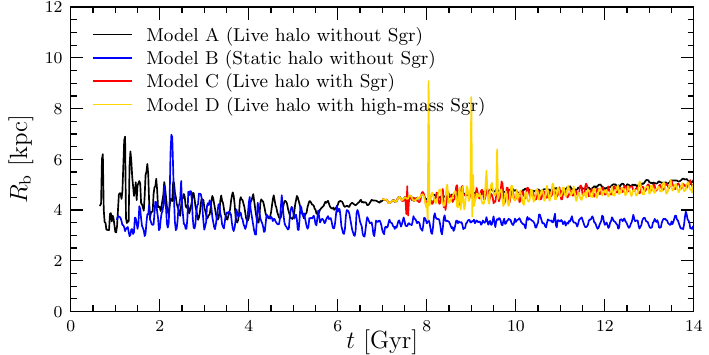}
    \includegraphics[width=8.5cm]{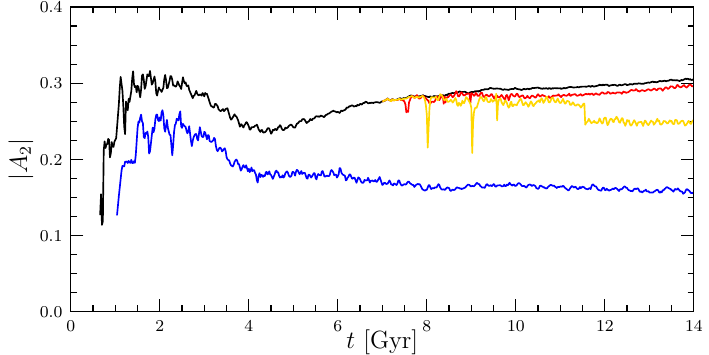}
    \includegraphics[width=8.5cm]{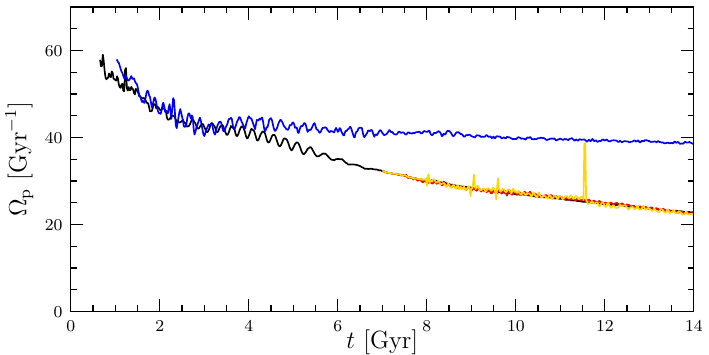}
    \includegraphics[width=8.5cm]{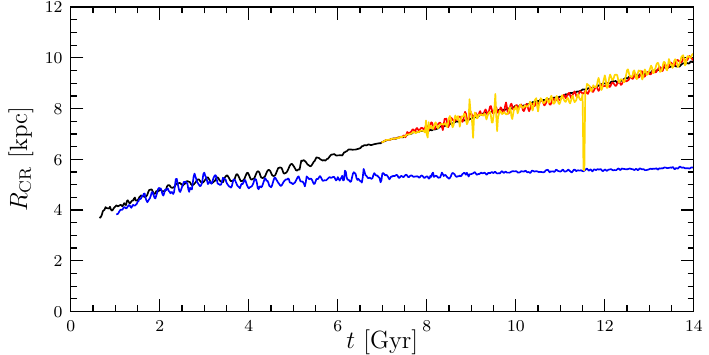}
    \includegraphics[width=8.5cm]{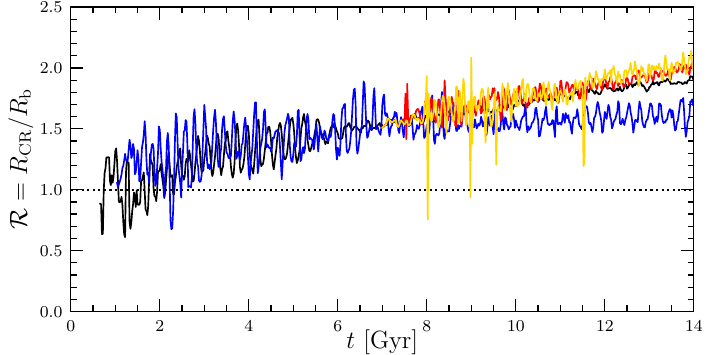}
    \caption{The bar length $\Rb$, amplitude $|A_2|$, pattern speed $\Omegap$, corotation radius $\RCR$, and the ratio $\mathcal{R}=\RCR/\Rb$ as a function of time.}
    \label{fig:bar_properties}
  \end{center}
\end{figure}

This Appendix presents the time evolution of the bar's properties in our simulations. We quantify the bar using the $m=2$ azimuthal Fourier coefficient of the stellar surface density $\Sigma$:
\begin{align}
  A_2(R) = \frac{\hat{\Sigma}_2(R)}{\hat{\Sigma}_0(R)}, 
  \label{eq:A_2}
\end{align}
where
\begin{align}
  \hat{\Sigma}_m(R) &= \frac{1}{2\pi}\int \drm \varphi \,\Sigma(\varphi,R) \e^{-\mi m \varphi}. 
  \label{eq:Sigma_m}
\end{align}
In practice, we evaluate this in radial bins of finite width $\Delta R=0.2 \kpc$. Thus,
\begin{align}
  A_2(R) = \frac{\sum_i \mu_i \e^{-\mi 2 \varphi_i}}{\sum_i \mu_i},
  \label{eq:A_2_discrete}
\end{align}
where the summation is carried out over particles with mass $\mu_i$, angle $\varphi_i$, and radius $R_i \in [R-\Delta R/2,R+\Delta R/2]$. We define the bar length $\Rb$ as the radius at which the amplitude $|A_2|$ falls below half of its maximum value $|A_{2,\mathrm{max}}|$. We then recompute the Fourier coefficient of $\Sigma$ over the radial interval $[0,\Rb]$ and take its magnitude and argument as the bar amplitude and phase, respectively. The bar pattern speed is obtained by taking finite differences of the bar phase in time.

Fig.\,\ref{fig:bar_properties} shows, from top to bottom, the time evolution of the bar length, amplitude, pattern speed, corotation radius $\RCR$, and the ratio $\mathcal{R}\equiv\RCR/\Rb$, for models A (black), B (blue), C (red), and D (orange). Consistent with previous studies, the bar in a live dark halo forms faster \citep[e.g.][]{Athanassoula2002BarGrowth,Frosst2024active} and later spins down and grows in both length and amplitude \citep[e.g.][]{debattista2000constraints,Dubinski2009Anatomy,Fujii2019DryGalaxy}, while the bar in the static halo keeps its shape and speed almost constant (although it spins down a little by transferring angular momentum to the outer stellar disk). The passage of Sagittarius systematically shortens and weakens the bar, while leaving the pattern speed largely unchanged. This contrasts with the study by \cite{Kodama2026Galaxy}, which finds that past interactions with a massive satellite, even prior to bar formation, can promote the bar's secular evolution. A key difference is that their simulation adopts a model in which the bar enters a metastable state \citep{Sellwood2006Metastability}, making it susceptible to small fluctuations in the potential. In addition, while we adopt a satellite on a polar orbit, they consider a satellite on an in-plane orbit, resulting in a stronger in-plane perturbation.

The bottom panel of Fig.\,\ref{fig:bar_properties} shows the evolution of the ratio $\mathcal{R}\equiv\RCR/\Rb$, which has been widely used as a diagnostic of bar evolution. Early simulations of non-growing disks \citep[e.g.][]{debattista2000constraints} found that $\mathcal{R}$ increases as the bar slows, a trend that we also confirm in our non-growing disk simulations. In contrast, simulations that include realistic disk growth find a more constant $\mathcal{R}$ \citep{aumer2015origin,Dehnen2023Measuring}, as bars form short and later elongate substantially in line with the expanding corotation radius. A similar coevolution of $\RCR$ and $\Rb$ is seen in the IllustrisTNG50 cosmological simulation \citep{Semczuk2024Pattern}.

\section{Computation of the libration action}
\label{sec:computation_of_Jl}

\begin{figure}
  \begin{center}
    \includegraphics[width=8.5cm]{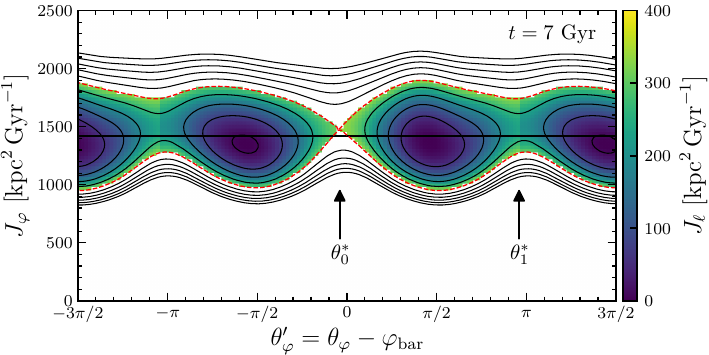}
    \caption{Libration action $\Jl$ computed over the slow angle-action space for Model A at $t=7 \Gyr$. Black curves mark iso-contours of the averaged Hamiltonian $\bH$ (Fig.\,\ref{fig:xy_f_Jphi0_H}). The red dashed curve marks the separatrix, defined by the lower saddle point ($\theta_0^\ast$) of $\bH$. Trapped regions where contours enclose both resonant islands are split at the higher saddle point ($\theta_1^\ast$) of $\bH$. The libration action in this region is calculated separately for each island using the divided contours.}
    \label{fig:Jl_thppJphi}
  \end{center}
\end{figure}

Here we describe the procedure to compute the libration action $\Jl$ (equation \ref{eq:Jl}) from the map of the fast-angle-averaged Hamiltonian $\bH$ (equation \ref{eq:aveH}). 

The libration action is given by the phase-space area enclosed by the contours of constant $\bH$. Our first task is therefore to identify the iso-contours of $\bH$ from its values precomputed on a discrete grid in $(\thetaphi', \Jphi)$ for a given set of fast actions $\vJf = (\Jr, \Jz)$. To this end, we use the marching squares algorithm \citep{lorensen1998marching} as implemented in \texttt{scikit-image} \citep{scikit-image}. We then compute the area enclosed by each contour using the shoelace formula \citep{braden1986surveyor},
\begin{align}
  \Jl = \frac{1}{4\pi} \left| \sum_{i=1}^{N} \left( \thetaphi'^{\,i}\Jphi^{\,i+1} - \thetaphi'^{\,i+1}\Jphi^{\,i} \right) \right|,
  \label{eq:Jl_shoelace}
\end{align}
where $N$ is the number of points along the contour, and $X^{N+1} \equiv X^{1}$ so that the contour is closed.

Figure\,\ref{fig:Jl_thppJphi} shows an example of $\Jl$ computed with this method for Model~A at $t=7\Gyr$. The black curves show the iso-contours of $\bH$ (not $\Jl$). The Hamiltonian has two saddle points, whose angle we denote as $\theta^\ast_0$ and $\theta^\ast_1$. The separatrix is defined as the contour that passes through the lower of the two, $\theta_0^\ast$, as marked by the red dashed curve. With this definition, contours near the separatrix may enclose both resonant islands. If one were to compute the total area enclosed by these contours, $\Jl$ would be approximately twice as large as those that only enclose one of the islands. To avoid such a discontinuity, we split these contours at the higher saddle point, $\theta_1^\ast$, and compute the libration action for each island separately using the split contours. This ensures that $\Jl$ varies smoothly across the contours of $\bH$ and remains a physically meaningful measure of the trapped volume around each resonant island.

Calculating the libration action for each individual particle is computationally demanding, as it requires generating a map of $\bH$ over the slow/azimuthal plane for every particle, each of which has different fast actions $\vJf = (\Jr, \Jz)$. To reduce the computational cost, we precompute $\bH$ on a $100 \times 100$ grid in $(\thetaphi', \Jphi) \in [0,2\pi] \times [0,3000]\,\kpckpcGyr$ for $\vJf$ sampled on a $5 \times 5$ grid spanning $[0,100] \times [0,100]\,\kpckpcGyr$. This results in a total of $2.5 \times 10^5$ grid points for each snapshot. We then precompute $\Jl$ over the same grid and obtain $\Jl$ of each particle by linear interpolation.

\section{Impact of asymmetry of the Hamiltonian}
\label{sec:asymmetry_Hamiltonian}

\begin{figure*}
  \begin{center}  
    \includegraphics[width=18.0cm]{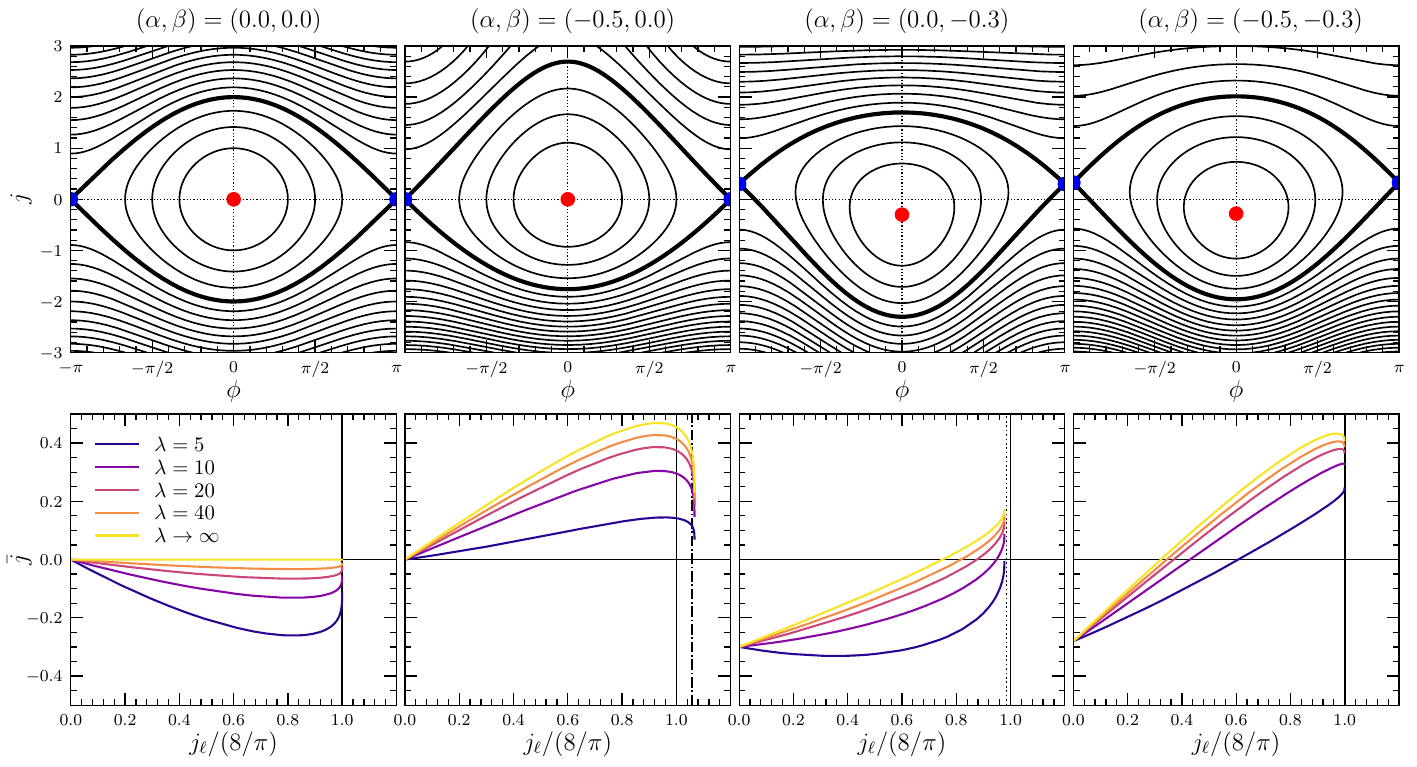}
    \caption{Impact of the asymmetry in the Hamiltonian on the tree-ring structure of the resonance. Top row: Level curves of the Hamiltonian (equation \ref{eq:aveH_dimensionless}) for different combinations of the asymmetry parameters $(\alpha,\beta)$ (equation \ref{eq:alph_beta}) denoted on the top. Red and blue dots mark the stable and unstable fixed points, respectively. Bottom row: Correlation between the libration action $\jl$ and the mean slow action $\bar{j}$, for five different values of the dimensionless scale length $\lambda$ of the initial distribution function $f_0 \propto \exp(-j/\lambda)$. The dot-dashed line (\ref{eq:jlsep_b0}) and dotted line (\ref{eq:jlsep_a0}) mark the analytical prediction for the maximum $\jl$ attained at the separatrix. In a symmetric Hamiltonian $(\alpha,\beta)=(0,0)$ (leftmost panel), the correlation is predominantly negative, because the phase-space density is higher at lower $j$. In an asymmetric Hamiltonian, however, the correlation becomes positive because (i) asymmetry in the unperturbed Hamiltonian $(\alpha \neq 0)$ shifts the contours toward larger $j$ while leaving the fixed points unchanged, and (ii) asymmetry in the bar potential $(\beta \neq 0)$ shifts the stable fixed point (red) downward and the unstable fixed points (blue) upward. Both effects result in $\bar{j}$ increasing with $\jl$.}
    \label{fig:H_Jl_mJp}
  \end{center}
\end{figure*}

This Appendix discusses the origin of the weak positive correlation between the stars' libration action $\Jl$ and their initial angular momentum $J_{\varphi 0}$ found in Model B, in which the bar maintains a constant pattern speed. If the Hamiltonian were perfectly symmetric about the resonance and the resonance were stationary, we would instead expect a negative correlation, because the initial phase-space density has a negative gradient in $\Jphi$: there are initially more particles below the resonance than above it, so the mean $\Jphi$ along curves of constant $\Jl$ would be smaller than the value at the resonance center ($\Jl=0$). The observed positive correlation must therefore originate from the asymmetric structure of the Hamiltonian.

To see how asymmetries in the Hamiltonian can generate a pseudo tree-ring structure, we analyze the averaged Hamiltonian $\bH$ introduced in equation~(\ref{eq:aveH}), considering only the bar's quadrupole component ($m=2$):
\begin{align}
  \bH(\thetas,\Js) = H_0(\Js) - \Nphi \Omegap \Js + 2 |\hPhi_1(\Js)| \cos(\thetas - \thetasres),
  \label{eq:aveH_quadrupole}
\end{align}
where $(\thetas,\Js)=(\Nphi\thetaphi',\Jphi/\Nphi)$ are the slow angle-action variables associated with the corotation resonance, and we have omitted reference to the fast actions $\vJf$, which are constants.
The quantity $|\hPhi_1|$ and $\thetasres$ are the amplitude and phase of the $m=2$ Fourier coefficient.
We assume throughout that the bar rotates steadily and does not grow.

The structure of $\bH$ near a resonance is commonly understood by expanding it around the resonance up to second order in $H_0$ and zeroth order in the perturbation,
\begin{align}
  \bH(\thetas,\Js) \simeq \frac{1}{2} H^{(2)}_0 (\Js - \Jsres)^2 + 2 |\hPhi_1|^{(0)} \cos(\thetas - \thetasres),
  \label{eq:aveH_pendulum}
\end{align}
where $X^{(n)}$ denotes the $n$th derivative with respect to $\Js$ evaluated at the resonance $\Jsres$.
The resulting Hamiltonian takes the form of a pendulum and is evidently symmetric about the resonance \citep[e.g.][]{lichtenberg1992regular}.

To assess the impact of the asymmetry in $\bH$, we extend the expansion to the next order:
\begin{align}
  &\bH(\thetas,\Js) \simeq \frac{1}{2} H^{(2)}_0(\Js - \Jsres)^2 + \frac{1}{6} H^{(3)}_0(\Js - \Jsres)^3 \nonumber \\
  &+ 2 \left[|\hPhi_1|^{(0)} + |\hPhi_1|^{(1)} (\Js - \Jsres)\right] \cos(\thetas - \thetasres).
  \label{eq:aveH_asymmetry}
\end{align}
For convenience, we now introduce the following dimensionless variables:
\begin{align}
  &\tau = t \omega_0, ~~\phi \equiv \thetas - \thetasres, ~~ j \equiv \frac{\Js - \Jsres}{J_0}, 
  \label{eq:dimensionless_variables}\\
  &\alpha \equiv \frac{H^{(3)}_0 J_0}{H^{(2)}_0}, ~~ \beta \equiv \frac{|\hPhi_1|^{(1)} J_0}{|\hPhi_1|^{(0)}},
  \label{eq:alph_beta}
\end{align}
where 
\begin{align}
  \omega_0 = \sqrt{2|\hPhi_1|^{(0)} |H^{(2)}_0|}, ~~ J_0 \equiv \sqrt{2|\hPhi_1|^{(0)} / |H^{(2)}_0|}, 
  \label{eq:J0}
\end{align}
denote the characteristic libration frequency and resonance width, respectively. The parameter $\alpha$ measures the asymmetry of the unperturbed Hamiltonian, while $\beta$ quantifies the asymmetry of the bar potential. Assuming $H^{(2)}_0<0$ and rescaling the Hamiltonian by $2|\hPhi_1|^{(0)}$, we obtain
\begin{align}
  \mathcal{H}(\phi,j) = - \frac{1}{2} j^2 - \frac{\alpha}{6} j^3 + \left(1 + \beta j\right) \cos \phi.
  \label{eq:aveH_dimensionless}
\end{align}
Note that $(\phi,j)$ constitute a canonical pair, since the equations of motion take the standard Hamiltonian form, $(\dot{\phi},\dot{j})=(\pd \mathcal{H}/\pd j, -\pd \mathcal{H}/\pd \phi,)$, with respect to the rescaled time $\tau$ (\ref{eq:dimensionless_variables}). The stable and unstable fixed points of $\mathcal{H}$ occur at
\begin{align}
  (\phi^{\ast}_{\rm s},j^{\ast}_{\rm s}) &= \left(0, \frac{-1 + \sqrt{1+2\alpha \beta}}{\alpha}\right), \\
  (\phi^{\ast}_{\rm u},j^{\ast}_{\rm u}) &= \left(\pm \pi, \frac{-1 + \sqrt{1-2\alpha \beta}}{\alpha}\right).
  \label{eq:fixed_point}
\end{align}

For reference, typical values of the parameters in Model B at $t \sim 4 \Gyr$ are
\begin{equation}
  \begin{aligned}
    H^{(2)}_0 &\sim - 0.17  \kpc^{-2} \\
    H^{(3)}_0 &\sim 9.1 \times 10^{-4} \Gyr\kpc^{-4} \\
    |\hPhi_1|^{(0)} &\sim 840 \kpc^{2}\Gyr^{-2} \\
    |\hPhi_1|^{(1)} &\sim - 2.7 \Gyr^{-1}
  \end{aligned}
  \xrightarrow{\quad}
  ~
  \begin{aligned}
    \omega_0 &\sim 17 \Gyr^{-1} \\
    J_0 &\sim 100 \kpckpcGyr \\
    \alpha &\sim -0.5 \\
    \beta &\sim -0.3
  \end{aligned}
  \label{eq:coefficients_values}
\end{equation}

The top row of Fig.~\ref{fig:H_Jl_mJp} shows the contours of Hamiltonian for several pairs of $(\alpha,\beta)$. The leftmost panel corresponds to the standard pendulum Hamiltonian, $(\alpha,\beta)=(0,0)$, which is manifestly symmetric about the resonance, $j=0$. The second panel from the left shows the case $(\alpha,\beta)=(-0.5,0)$, illustrating the effect of asymmetry in $H_0$ alone. The contours become more widely spaced above the resonance and more closely spaced below it. The third panel from the left presents the case $(\alpha,\beta)=(0,-0.3)$, highlighting the asymmetry introduced by $|\hPhi_1|$. The primary effect here is the displacement of the fixed points: the resonance center (red dot) shifts downward, whereas the saddle points (blue dots) shift upward, each by an amount $|\beta|$ \citep[see also][]{Kaasalainen1994hamiltonian}. Finally, the rightmost panel shows $(\alpha,\beta)=(-0.5,-0.3)$, which combines both effects.

We now calculate the libration action $\jl$ and the mean $j$ along these contours. The dimensionless libration action is
\begin{align}
  \jl \equiv \frac{\Jl}{J_0} = \frac{1}{2\pi} \oint \drm \phi j(\mathcal{H},\phi).
  \label{eq:jl}
\end{align}
Its maximum value, attained at the separatrix, is 
\begin{align}
  \jlsep \equiv \frac{1}{2\pi} \oint \drm \phi j[\mathcal{H}(\phi^{\ast}_{\rm u},j^{\ast}_{\rm u}), \phi].
  \label{eq:jlsep}
\end{align}
For $\alpha=0$, an exact analytic expression can be derived:
\begin{align}
  \jlsep &= \frac{4}{\pi} \left( \frac{\sin^{-1}(\beta)}{\beta} + \sqrt{1 - \beta^2} \right) \nonumber \\
  &\simeq \frac{8}{\pi}\left[1 - \frac{\beta^2}{6} - \frac{\beta^4}{40} + \mathcal{O}(\beta^6)\right].
  \label{eq:jlsep_a0}
\end{align}
The equation shows that asymmetry in the bar potential---irrespective of its sign---works to decrease the trapped phase-space volume. For $\beta=0$, we expand the equation for the trajectory of the separatrix, $j^\ast=j[\mathcal{H}(\phi^{\ast}_{\rm u},j^{\ast}_{\rm u}), \phi]$, as a power series in $\alpha$
\begin{align}
  j^\ast = j^\ast_0 + j^\ast_1 \alpha + j^\ast_2 \alpha^2 + j^\ast_3 \alpha^3 + \dots
  \label{eq:jlsep_a_expansion}
\end{align}
and solve order by order. Substituting to (\ref{eq:jlsep}), we obtain
\begin{align}
  \jlsep \simeq \frac{8}{\pi}\left[1 + \frac{5}{27}\alpha^2 + \frac{77}{405}\alpha^4 + \mathcal{O}(\alpha^6)\right].
  \label{eq:jlsep_b0}
\end{align}
Hence, in contrast to the effect of $\beta$, asymmetry in the unperturbed Hamiltonian increases the trapped volume.

The mean (orbit-averaged) $j$ is 
\begin{align}
  \bar{j}(\jl) \equiv \frac{\int_0^{2\pi} \drm \phil f_0(j) j(\phil,\jl)}{\int_0^{2\pi} \drm \phil f_0(j)},
  \label{eq:meanj}
\end{align}
where $\phil$ is the libration angle conjugate to $\jl$, and $f_0(j)$ is the initial distribution function. For simplicity, we model $f_0$ as exponential in $j$, i.e. $f_0 \propto \exp(-j/\lambda)$, where $\lambda$ is the dimensionless scale length. In our simulations, the initial distribution has $\lambda \sim 5$. The limit $\lambda \rightarrow \infty$ corresponds to a uniform distribution. In practice, the orbit average in equation~(\ref{eq:meanj}) is numerically computed as a time average over one libration period $\Tl \equiv \oint \drm \phi / \dot{\phi}$.

The bottom row of Fig.~\ref{fig:H_Jl_mJp} plots $\bar{j}$ against $\jl$ for various values of $\lambda$. In the standard pendulum model (leftmost panel), $\bar{j}$ declines with $\jl$ because $f_0'<0$, as predicted. Near the separatrix, however, $\bar{j}$ rises back toward the origin because stars linger near the saddle points (blue dots in the upper panel), which enhances the weight at $j=0$. As $\lambda$ increases, the initial distribution becomes flatter, and consequently $\bar{j}$ depends more weakly on $\jl$. In the limit $\lambda \rightarrow \infty$, $\bar{j}$ becomes independent of $\jl$.

For the case $(\alpha,\beta)=(-0.5,0)$ (second panel from the left), we now observe a significant positive correlation, except in the close vicinity of the separatrix for the same reason discussed above. This positive correlation arises because the phase-space volume within a given interval of $\jl$ is larger at positive $j$ than at negative $j$. This asymmetry in phase-space volume outweighs the asymmetry in phase-space density, with the result that $\bar{j}$ increases with $\jl$. We also confirm that increasing $\alpha$ enlarges the resonant volume, in agreement with our analytic prediction (\ref{eq:jlsep_b0}) marked by dot-dashed line.

For the case $(\alpha,\beta)=(0,-0.3)$ (third panel from the left), a positive correlation again appears, but for a qualitatively different reason. Here, the stable (unstable) fixed points are shifted down (up), so the mean $j$ at the endpoints are correspondingly displaced to $\bar{j}=\beta=-0.3$ at $\jl=0$ and $\bar{j}=-\beta=0.3$ at $\jl=\jlsep$, regardless of $\lambda$. These changes in the endpoints enforce a net positive slope between them. Again, we confirm that increasing $\beta$ shrinks the resonant volume (\ref{eq:jlsep_a0}, dotted line).

Finally, the case $(\alpha,\beta)=(-0.5,-0.3)$ (rightmost panel), which most closely resembles our simulation (Model B), exhibits a positive correlation due to the combined effect of the asymmetry in both the unperturbed Hamiltonian and the bar potential. The total variation in $\bar{j}$ from the resonance center to the separatrix is $\Delta \bar{j} \sim 0.5$, which corresponds to an increase in angular momentum of $\Delta\bar{J}_{\varphi} = \Nphi \Delta \bar{j} J_0 \sim 100\kpckpcGyr$, consistent with the increase measured in the simulation (Fig.\,\ref{fig:Jp0_Jl}, Model B).

\section{Recovering the bar's pattern speed from the tree-ring structure}
\label{sec:infer_wp}

\begin{figure}
  \begin{center}  
    \includegraphics[width=8.5cm]{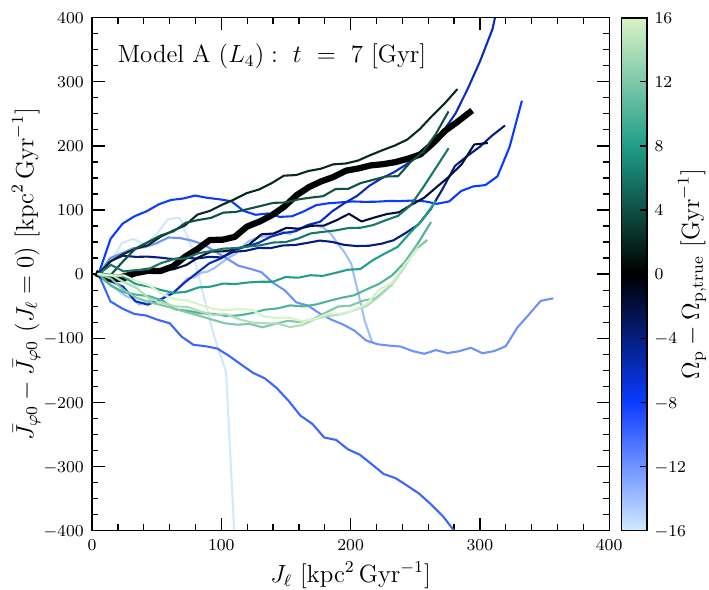}
    \includegraphics[width=8.5cm]{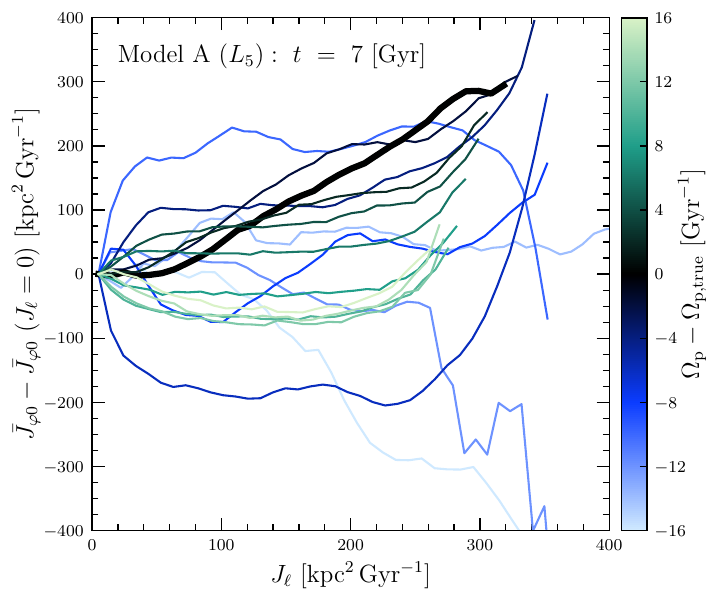}
    \caption{Correlation between the initial angular momentum $J_{\varphi 0}$ and the normalized angle-averaged Hamiltonian $h$ for different assumed values of bar pattern speed. The top and bottom figures show the results for the two resonant islands surrounding the Lagrange points $L_4$ and $L_5$. A positive correlation appears only when the pattern speed is close to the true value (black).}
    \label{fig:Jp0_Jl_wp}
  \end{center}
\end{figure}
\begin{figure}
  \begin{center}  
    \includegraphics[width=8.5cm]{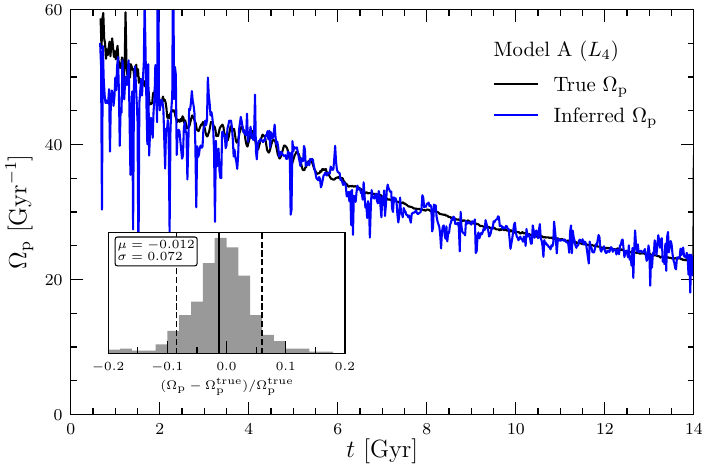}
    \includegraphics[width=8.5cm]{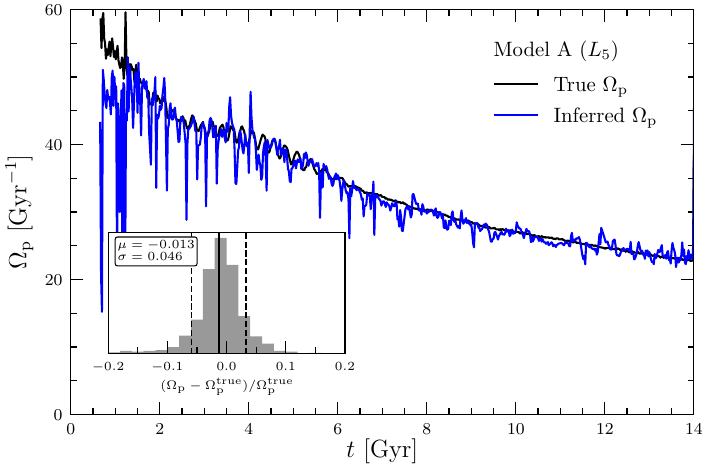}
    \caption{Bar pattern speed of Model~A inferred by demanding a tree-ring structure inside the corotation resonance. As in Fig.~\ref{fig:Jp0_Jl_wp}, the top and bottom panels correspond to the resonant islands around $L_4$ and $L_5$, respectively. The inset shows the distribution of the fractional residuals between the inferred and true values for $t > 2 \Gyr$.}
    \label{fig:infer_wp}
  \end{center}
\end{figure}

In this appendix, we show how the tree-ring structure of the bar's resonance can be used to infer the bar's instantaneous pattern speed $\Omegap$. The identification of the tree-ring structure requires knowledge of the exact location of the resonance set by $\Omegap$. If an incorrect pattern speed is adopted, the contours of $\Jl$ become offset from those of $J_{\varphi 0}$, resulting in a loss of coherence between the two quantities. We illustrate this in Figure\,\ref{fig:Jp0_Jl_wp}, which shows the relation between $\Jl$ and $J_{\varphi 0}$ for a range of pattern speeds. When the correct pattern speed is adopted (thick black), $\Jl$ and $J_{\varphi 0}$ exhibit a tight, coherent relation, as expected. However, when the pattern speed deviates from the true value, this coherence degrades. 

This sensitivity implies that we may infer the correct pattern speed by searching for the value that maximizes the correlation between $\Jl$ and $J_{\varphi 0}$. \cite{Chiba2021TreeRing} applied this idea to measure the present-day pattern speed of the Galactic bar, finding good agreement with independent estimates based on direct kinematic modelling of stars in the bar/bulge region \citep{Clarke2021ViracGaia,Leung2023measurement,Zhang2024Kinematics}. Here, we adopt the same approach and assess how accurately the true pattern speed can be recovered from the simulated data.

Similar to \cite{Chiba2021TreeRing}, we treat the local gradient $x_i \equiv \drm J_{\varphi 0}/ \drm \Jl|_i = (J_{\varphi 0}^i - J_{\varphi 0}^{i-1}) / \Delta \Jl$ between adjacent bins in $\Jl$ as a Gaussian random variable with mean $\mu_i$ and uncertainty $\sigma_i$. The probability that the gradient is positive $(x_i > 0)$ is
\begin{align}
  \mathcal{L}_i(\Omegap) = \frac{1}{2}\left[1 + {\rm erf}\left(\frac{\mu_i}{\sqrt{2}\sigma_i}\right)\right].
  \label{eq:likelihood}
\end{align}
The total likelihood that $J_{\varphi 0}$ increases monotonically with $\Jl$ is then obtained by multiplying the contributions from all bins,
\begin{align}
  \mathcal{L}(\Omegap) = \prod_{i}^{N} \mathcal{L}_i(\Omegap) = \prod_{i}^{N}  \frac{1}{2}\left[1 + {\rm erf}\left(\frac{\mu_i}{\sqrt{2}\sigma_i}\right)\right].
  \label{eq:total_likelihood}
\end{align}

Figure\,\ref{fig:infer_wp} compares the true pattern speed (black) with the inferred values (blue) for the two resonant islands around $L_4$ (top) and $L_5$ (bottom). The inference exhibits substantial noise, particularly at early times when the spiral structure is strong, but the noise lessens as the bar becomes dominant. Nevertheless, the method recovers the true pattern speed with no significant systematic bias. As shown in the inset panel, the fractional residuals have a mean negative offset of $\sim 1\%$ and a standard deviation of $\sim 5-7 \%$. This suggests that the value measured in the Milky Way using this method, $35.5 \pm 0.8 \kmskpc$ \citep{Chiba2021TreeRing}, may be subject to a systematic bias of $\sim 0.4 \kmskpc$ and may carry an additional uncertainty of $\sim 2.5 \kmskpc$. This level of bias and uncertainty remains consistent with other independent measurements \citep{Clarke2021ViracGaia,Leung2023measurement,Zhang2024Kinematics}.

%% For this sample we use BibTeX plus aasjournalv7.bst to generate the
%% the bibliography. The sample7.bib file was populated from ADS. To
%% get the citations to show in the compiled file do the following:
%%
%% pdflatex sample7.tex
%% bibtext sample7
%% pdflatex sample7.tex
%% pdflatex sample7.tex

\bibliography{./references}{}
\bibliographystyle{aasjournalv7}

%% This command is needed to show the entire author+affiliation list when
%% the collaboration and author truncation commands are used.  It has to
%% go at the end of the manuscript.
%\allauthors

%% Include this line if you are using the \added, \replaced, \deleted
%% commands to see a summary list of all changes at the end of the article.
%\listofchanges

\end{document}